\documentclass[sigconf]{acmart}
\usepackage{array}
\usepackage{float}
\usepackage{subcaption}
\usepackage[linesnumbered,ruled,vlined]{algorithm2e}
\usepackage{comment}
\AtBeginDocument{%
  }
\copyrightyear{2026}
\acmYear{2026}
\setcopyright{cc}
\setcctype{by}
\acmConference[CHI '26]{Proceedings of the 2026 CHI Conference on Human Factors in Computing Systems}{April 13--17, 2026}{Barcelona, Spain}
\acmBooktitle{Proceedings of the 2026 CHI Conference on Human Factors in Computing Systems (CHI '26), April 13--17, 2026, Barcelona, Spain}
\acmPrice{}
\acmDOI{10.1145/3772318.3790655}
\acmISBN{979-8-4007-2278-3/2026/04}

\begin{document}

\title{When Should Users Check? Modeling Confirmation Frequency in Multi-Step Agentic AI Tasks}
\author{Jieyu Zhou}
\affiliation{%
  \institution{Georgia Institute of Technology}
  \city{Atlanta}
  \state{Georgia}
  \country{USA}
}
\email{jzhou625@gatech.edu}

\author{Aryan Roy}
\affiliation{%
  \institution{Georgia Institute of Technology}
  \city{Atlanta}
  \state{Georgia}
  \country{USA}
}
\email{aroy389@gatech.edu}

\author{Sneh Gupta}
\affiliation{%
  \institution{Georgia Institute of Technology}
  \city{Atlanta}
  \state{Georgia}
  \country{USA}
}
\email{sgupta852@gatech.edu}

\author{Daniel Weitekamp}
\affiliation{%
  \institution{Georgia Institute of Technology}
  \city{Atlanta}
  \state{Georgia}
  \country{USA}
}
\email{dweitekamp3@gatech.edu}

\author{Christopher J. MacLellan}
\affiliation{%
  \institution{Georgia Institute of Technology}
  \city{Atlanta}
  \state{Georgia}
  \country{USA}
}
\email{cmaclell@gatech.edu}

\begin{teaserfigure}
  \centering
  \includegraphics[width=\textwidth]{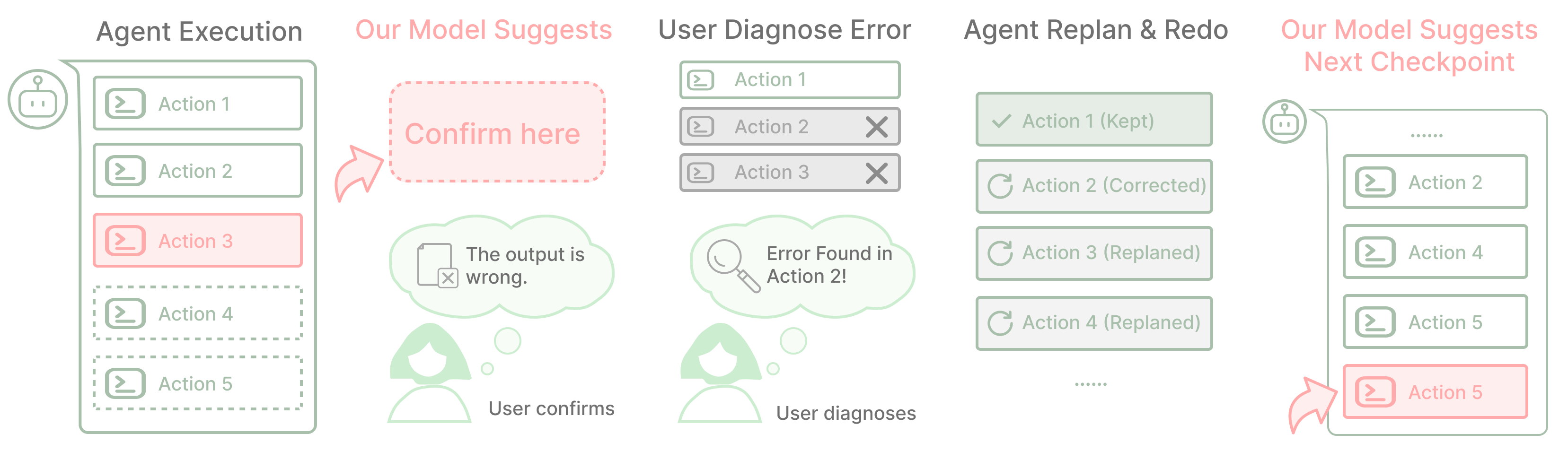}
  \caption{Confirmation–Diagnosis–Correction–Redo (CDCR) Pattern. Based on the expected time of user interaction under the CDCR pattern, our model decides whether the agent should proceed autonomously or prompt the user to step in and supervise.}
  \Description{Confirmation–Diagnosis–Correction–Redo (CDCR)}
  \label{fig:CDCR}
\end{teaserfigure}

\begin{abstract}
Existing AI agents typically execute multi-step tasks autonomously and only allow user confirmation at the end. During execution, users have little control, making the confirm-at-end approach brittle: a single error can cascade and force a complete restart. Confirming every step avoids such failures, but imposes tedious overhead. Balancing excessive interruptions against costly rollbacks remains an open challenge. We address this problem by modeling confirmation as a minimum time scheduling problem. We conducted a formative study with eight participants, which revealed a recurring \textit{Confirmation–Diagnosis–Correction–Redo (CDCR)} pattern in how users monitor errors. Based on this pattern, we developed a decision-theoretic model to determine time-efficient confirmation point placement. We then evaluated our approach using a within-subjects study where 48 participants monitored AI agents and repaired their mistakes while executing tasks. Results show that 81\% of participants preferred our intermediate confirmation approach over the confirm-at-end approach used by existing systems, and task completion time was reduced by 13.54\%. 
\end{abstract}
\maketitle


\section{Introduction}
With the advancement of Artificial Intelligence (AI), Agentic AI has emerged as a concept describing agents that assist humans in carrying out complex tasks \cite{durante2024agent, pallagani2024prospects, huang2024understanding, wang2024survey, yao2023react, sumers2023cognitive, schick2023toolformer, wei2022chain}. These AI agents have been deployed across diverse domains, including multi-modal devices \cite{rabbit2023, chang2024partnr}, web-based systems \cite{huq2023s,li2023using, li2017sugilite}, and game controls \cite{wang2023voyager,wang2023describe,lawley2023val}. Their rapid proliferation has reignited a long-standing HCI debate \cite{horvitz1999principles, shneiderman1997direct, lieberman1997autonomous, maes1995agents}: should agents operate under direct human supervision and control or should they operate in a mostly autonomous fashion?

Most commercial AI agents today prioritize end-to-end autonomy, and rarely proactively ask users to confirm intermediate steps \cite{openai2025agent, anthropic2024claude, manus2024}. While some agents allow users to take control or interrupt execution \cite{openai2025agent, qin2025ui}, these mechanisms are reactive: they are only initiated when users halt execution upon noticing something is wrong, or when the agent halts to request missing information. Without proactive reminders, users often miss the window for timely intervention. As a result, errors frequently go unnoticed until the end of the trajectory. This is problematic because current agents remain highly error-prone, achieving only ~30\% accuracy on multi-step benchmarks \cite{ma2024spreadsheetbench, xie2024osworld, xu2024theagentcompanybenchmarkingllmagents}, largely due to their limited capacity for generalizable reasoning and planning \cite{stechly2024chain, valmeekam2024planning}. Errors are especially costly in long-horizon agentic tasks, which often span 30 steps and 10–30 minutes of execution \cite{hu2024dawn, bonatti2024windows}. A single mistake can cascade into complete task failure, forcing users to redo the entire process—doubling execution time and, through repeated API calls, adding both monetary costs \cite{kapoor2024ai, feng2024cocoa, zhang2025leveraging} and increasing the agent's carbon footprint \cite{patterson2022carbon, luccioni2023estimating}. This motivates the need to determine when agents should proactively elicit user confirmation to prevent cascading failures.

The timing of such inspections is essentially a tradeoff between autonomy and control. At one extreme, too much user control (e.g., the user confirming after every step) prevents error propagation but imposes substantial interaction overhead, reducing efficiency to the level of manual execution or even worse. At the other extreme, too much automation (e.g., the user only confirming at the end) reduces the burden on the user but increases the risks of errors and costs associated with re-executing the tasks. The challenge is deciding when user checks should occur to balance the efficiency of automation with the safeguards of user control.

Prior HCI research has largely focused on \textbf{how} to provide user control, through visualizing the reasoning process, editing execution results, and providing error feedback \cite{arawjo2024chainforge, epperson2025interactive, kazemitabaar2024improving, feng2024cocoa, lam2022more, ashktorab2019resilient, li2020multi, levy2021assessing}. Some recent systems also consider \textbf{when} assistance should occur in user-led tasks, where the human is actively performing the activity and the agent decides when to offer help based on heuristic triggers or models of the user’s internal state \cite{pu2025promemassist, li2025satori, arakawa2024prism, pu2025assistance}. In contrast, we study the mirror case of agent-led workflows: an automated agent executes a multi-step task on the user’s behalf. The key question in this case is when the user should be brought back in to confirm. A notable exception is Horvitz’s seminal work on expected utility models for single-event decisions \cite{horvitz1999principles}, but these approaches do not extend to long-horizon tasks, where sequential dependencies and cumulative costs shape overall outcomes.  

In this paper, we address the open question of \textbf{when user checks should occur in long-horizon agentic tasks}. We first review representative agent benchmarks and deployed systems and present the results of a formative study we conducted with eight participants. The study revealed dissatisfaction with the confirm-at-end strategy used by existing agentic platforms. It also surfaced a recurring \textbf{Confirmation–Diagnosis–Correction–Redo (CDCR)} pattern employed by users to supervise agents. Building on these insights, we developed a decision-theoretic model that identifies periodic confirmation points that minimize total expected task completion time while ensuring task correctness. We directly validated our approach by comparing it to the confirm-at-end strategy using a within-subjects study with 48 participants under three domains. Results show that 81\% of participants (39 of 48) preferred our intermediate confirmation strategy over the confirm-at-end baseline. We also found that our strategy reduced the average task completion time by 13.54\% compared to the baseline. We further discuss parameter realism, present simulations that demonstrate the model's behavior under step-wise heterogeneity, and outline how our algorithm can be integrated to existing agent systems as a lightweight timing layer that triggers user-control hooks.

This paper makes four key contributions: 

1) A formative study identifying user expectations for alternative confirmation strategies compared to the current confirm-at-end approach, and revealing a recurring CDCR pattern;

2) A decision-theoretic model for checkpoint placement that minimizes user interaction while ensuring correct task execution;

3) An empirical evaluation with 48 participants demonstrating both strong user preference (81\%) for intermediate confirmation and a 13.54\% reduction in task completion time relative to the confirm-at-end baseline;

4) A discussion framing our model not only as a confirmation scheduling tool but also as a design probe to surface potential directions for user-supervised systems.

\section{Related Work}
HCI research has primarily examined how to give users control, while reliability engineering has modeled when to intervene. We unify these perspectives by using mathematical models grounded in user behavior to determine when human control should occur in long-horizon agentic tasks.

\subsection{Human Control in AI Agents}
The tension between system autonomy and direct user control has been a recurring theme in HCI for decades \cite{horvitz1999principles, shneiderman1997direct, lieberman1997autonomous, maes1995agents}. Current AI agents are highly automated, integrating context, tools, and memory to solve complex real-world tasks \cite{thrun2002probabilistic}. Yet these reasoning and tool-use systems are highly error-prone, yielding only ~30\% accuracy on 10-step tasks \cite{wei2022chain, dziri2023faith}. Errors propagate like a snowball—growing exponentially and causing downstream failures \cite{zhao2024retrieval, ji2023survey}. Agent self-correction is inconsistent and does not scale with task complexity \cite{kadavath2022language}. These challenges highlight the necessity of human control.

Along this mixed-initiative spectrum, recent work has examined user-led tasks in which the human is the primary actor and the system decides when to offer assistance. Systems such as Satori, PrISM-Observer, ProMemAssist, and proactive programming assistants model users’ activity or cognitive state and trigger help at predicted good moments \cite{li2025satori, arakawa2024prism, pu2025promemassist, pu2025assistance}. For example, agents intervene when a user is likely to reach a particular step, forget an action, or benefit from a reminder or suggestion. In these systems, the core timing question is when an assistant should interrupt or scaffold an ongoing user task.

On the agent-led side, HCI researchers have proposed a wide range of tools for human involvement in domain-specific workflows, particularly in programming and data analysis \cite{shao2024collaborative, das2024vime, zhao2024lightva, suh2024luminate, feng2024cocoa, lawley2023val, epperson2025interactive, kazemitabaar2024improving, chen2023miwa, zhou2025orchvis}. For instance, Cocoa builds on computational notebook interfaces to support co-planning and co-execution of research tasks \cite{feng2024cocoa}, AGDebugger enables pausing and editing agent behaviors in programming \cite{epperson2025interactive}, and Step-Phasewise decomposes complex analysis workflows for verification \cite{kazemitabaar2024improving}. In addition, research has expanded to general-purpose agents operating in open-ended environments. Systems like CowPilot \cite{huq2025cowpilot} and Magentic-UI \cite{mozannar2025magentic} have introduced mixed-initiative controls, enabling users to pause execution, reject plans, or approve specific steps. Similarly, Morae \cite{peng2025morae} and recent image creation agents \cite{hahnproactive} focus on detecting ambiguous decision points or maintaining explicit intent states to solicit user clarifications dynamically. These systems focus on how to give users control, but not when that control should be invoked in multi-step tasks. Our motivation is to provide a scheduling layer that triggers those existing hooks.

One useful way to study when interventions occur in agent-led workflows is through computational models of timing \cite{horvitz1999principles, cockburn2022probability, todi2021adapting, zhao2022bayesian}. For instance, \citet{horvitz1999principles} proposed expected-utility models for deciding whether to invoke services, and \citet{cockburn2022probability} showed that users’ probability-weighting biases shape when they engage or ignore assistive features. However, these works focus on single-event decisions and are not applicable to agentic AI, where tasks involve many interdependent steps. To capture the long-horizon nature of agent execution, we need to observe users’ sequential behaviors around errors and build mathematical models that identify the optimal moments for user confirmation from a global, system-level perspective.

\subsection{Error Prevention and Recovery Models}
Outside HCI, reliability engineering offers inspiration. Researchers have long developed mathematical models to optimize inspection intervals for complex systems such as railway tracks, drainage infrastructure, and manufacturing plants \cite{ten2013optimizing, iren2014cost, zhou2015preventive}. These studies show how inspection strategies affect whole-system performance, considering dependencies such as workforce allocation, spare parts, and interdependent subsystems \cite{dekker1997review, wang2002survey, de2020review}. The central challenge is to balance preventive maintenance (early inspections to prevent failures) with corrective maintenance (recovering after failures occur), minimizing long-run costs by trading off inspection labor and detection probabilities against repair and downtime losses \cite{assis2021dynamic, wang2012overview, barlow1960optimum, taghipour2012optimal}.
For AI agents, the analogy is clear: deciding when to confirm execution steps likewise balances preventive checks with corrective recovery. But unlike engineering, the relevant cost is not purely economic—it is shaped by user experience, including confirmation burden and error recovery effort.

HCI work on errors has largely focused on how to prevent or repair mistakes. Examples include encouraging users to issue longer or repeated commands for accuracy \cite{lam2022more}, designing conversations that make error recovery more resilient \cite{ashktorab2019resilient}, referring to third-party apps for additional information during repair \cite{li2020multi}, and studying whether error detection should be system- or user-initiated \cite{levy2021assessing}. These studies enrich our understanding of error handling, but they do not provide a mathematical account of when intervention should occur across multi-step tasks.

In summary, while prior HCI work has richly explored how to give users control and how to prevent or repair errors, it has rarely asked when such intervention should occur across long-horizon tasks. Similarly, reliability engineering provides rigorous models of inspection timing, but they optimize purely for economic cost in physical systems. Our contribution is to bridge these two traditions: we introduce a mathematical perspective on the \textit{when} question in agentic AI, grounded not in labor or repair costs, but in the dynamics of user experience—balancing confirmation burden against recovery effort. This shift reframes human–agent interaction as a problem of scheduling when to switch initiative between agents acting and users verifying and correcting, opening new space for both theoretical modeling and practical design of adaptive confirmation strategies.

\begin{figure*}
  \centering
  \includegraphics[width=1\linewidth]{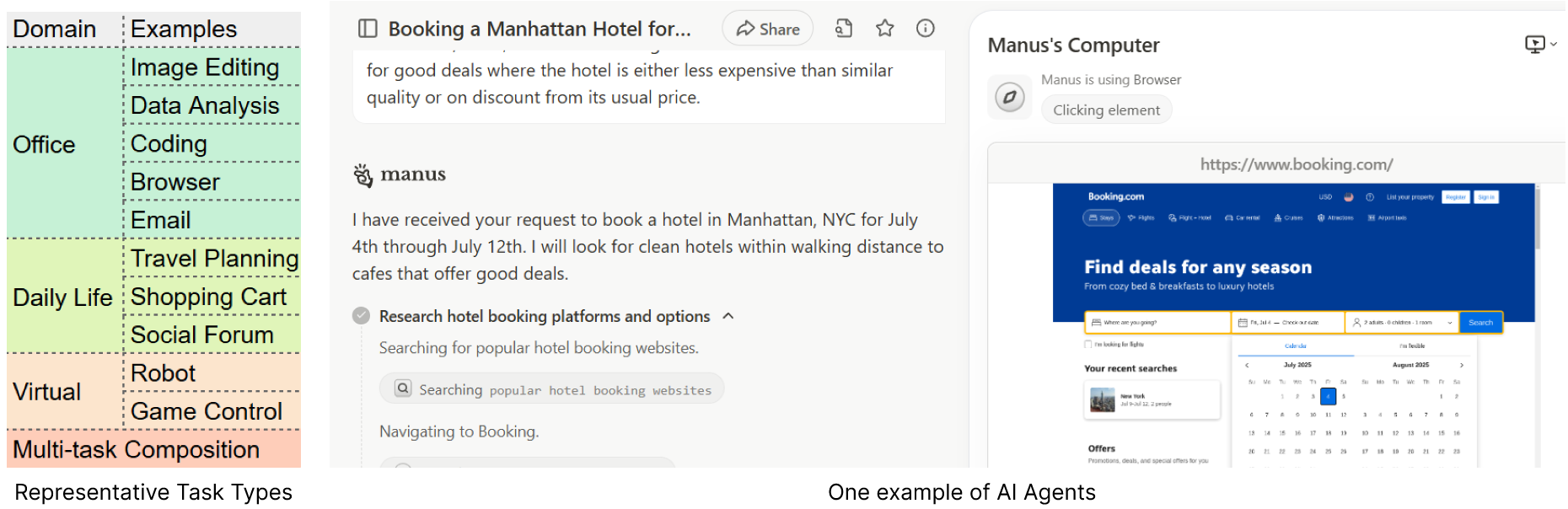}
  \caption{Representative Task Types and AI Agent Interface Example}
  \Description{Left: Task taxonomy generated from 12 representative benchmarks. Right: Example of an AI agent interface.}
  \label{fig:task type and interface}
\end{figure*}

\section{Grounding Confirmation Modeling in Agentic AI Contexts}
Before building the model, we conducted a formative study to address two research questions:
\begin{itemize}
    \item RQ1: How do users monitor AI agents during task execution? Is there any consistency in user behavior patterns across different task types and interface designs?
    \item RQ2: When should confirmation happen to support users in effectively monitoring AI agents?
\end{itemize}
To ensure that the observed user behaviors and the resulting modeling are grounded in realistic usage contexts, we first reviewed representative task benchmarks and existing AI agents (Section 3.1). Based on this review, we then designed a test environment that combines real-world tasks with deployed agentic systems to examine RQ1 and RQ2 (Section 3.2).

\subsection{Agentic AI Task Types and Existing Systems}
\subsubsection{Task Types}
To understand the landscape of agentic AI tasks and guide the design of our user study, we reviewed 12 benchmarks that cover a wide spectrum of real-world scenarios \cite{xie2024osworld, valmeekam2023planbench, xie2024travelplanner, zhou2023webarena, xu2024theagentcompanybenchmarkingllmagents, ma2024spreadsheetbench, jing2024dsbench, yao2022webshop, wei2025browsecomp, wang2024zsc, chang2024partnr}. These benchmarks are grounded in user-generated platforms such as StackOverflow, Reddit, and other public-facing systems, where users frequently ask how to complete complex or ambiguous tasks. Each benchmark contains around 300 pieces of task data. From these benchmarks, we identified four broad application domains—Office, Daily Life, Virtual Environment, and Mixed Workflow—each encompassing several representative task types (Figure~\ref{fig:task type and interface}, left). These domains provided the basis for selecting tasks in our formative study. Despite advances in LLMs, current models such as GPT-4 or Claude 3.5 still struggle to reliably complete long-horizon tasks. Reported success rates for benchmarks like OSWorld \cite{xie2024osworld}, SpreadsheetBench \cite{ma2024spreadsheetbench}, and TheAgentCompany \cite{xu2024theagentcompanybenchmarkingllmagents} often fall between 30\% and 60\%, with failure modes including incorrect tool use, misinterpreted instructions, or skipped steps. This relatively high error rate underscores the need for human-in-the-loop oversight, where users monitor and intervene during task execution.

\subsubsection{Existing Systems}
We tested tasks from the benchmarks mentioned above in nine agentic systems \cite{openai2025agent, anthropic2024claude, manus2024, flowith2024, crewai2024, layla2024, pokeeai2024, genspark2024, browseruse2024}. We also reviewed recent surveys on agent interface design \cite{luera2024survey, zheng2022ux, weisz2024design}. Typically, these agents operate within remote desktop environments, where the agent perceives its environment through periodic screenshots. The vision-language model (VLM) reasons about the current state and user instructions, and then performs primitive actions, such as mouse clicks and keyboard inputs. For instance, when the agent perceives a hotel booking website, it will enter the destination city on that website automatically, as shown in the middle of Figure~\ref{fig:task type and interface}. Over multiple steps, the agent decomposes a high-level goal into a sequence of fine-grained interface operations.

From the interaction perspective, we observed three common UI patterns across existing agents: (1) in some agents, each step is a clickable node linked to its corresponding screenshot \cite{manus2024, flowith2024, crewai2024, layla2024}; (2) in others, the agent’s activity or reasoning process is visually embedded within the screenshot, with a sliding progress bar to navigate through the log \cite{openai2025agent, anthropic2024claude, genspark2024}; and (3) some agents provide only screenshot playback without an activity log \cite{pokeeai2024, browseruse2024}. While these UI patterns allow users to inspect what the agent has done, confirmation mechanisms are exceedingly rare, given the low accuracy rates of current agents and the high cost of errors.  Most systems execute entire task sequences autonomously, while some can pause to let users take control \cite{openai2025agent, qin2025ui}. However, this pause mechanism is passive and reactive: the pause occurs only because the agent cannot proceed (e.g., missing payment credentials or login details), not because it seeks confirmation from users whether its execution so far is correct. As a result, reactive control provides no guidance on when users should check the agent’s progress, because it is entirely user-initiated. Consequently, most errors (e.g., clicking the wrong UI element or failing to load a page) are not detected in real time, and the agent either continues executing incorrectly or stops without recovery \cite{kadavath2022language}. These gaps motivated us to explore how to schedule intermediate confirmations in ways that ensure timely user intervention without introducing unnecessary overhead.

\subsection{Designing the Formative Study Environment}
Drawing from prior benchmarks and real-world agent deployments, we selected representative tasks to cover a broad spectrum of domains. By using real-world tasks and existing systems, we aim to replicate users’ authentic experiences, ensuring that subsequent model development would be grounded in realistic usage contexts.

\subsubsection{Environment Setup}
We selected five tasks to represent a broad spectrum of domains, including everyday activities, office workflows, and virtual environments. To cover the three UI patterns described above, we selected one representative agent for each pattern and randomly assigned tasks to these agents:
\begin{itemize}
    \item Agent 1 Manus \cite{manus2024} (node-based navigation) — (1) Travel planning: booking hotels, flights, taxis, and restaurants; (2) Shopping cart: adding ingredients for making pizza to an online shopping cart.
    \item Agent 2 ChatGPT Agent \cite{openai2025agent} (embedded sequential navigation) — (3) Document processing: converting scanned invoices into a spreadsheet and emailing it to the reimbursement office; (4) Image editing: removing unrelated objects, changing backgrounds, and replacing objects in an image.
    \item Agent 3 Browser Use \cite{browser_use2024} (screenshot-only playback) — (5) Video game play: navigating through different terrains, boiling ingredients, and plating meals in the game Overcooked.
\end{itemize}
The Overcooked task runs in a custom game environment that requires direct programmatic control and access to internal state. In our setup, only Browser Use could reliably operate in this environment, so the Overcooked task was always paired with Agent~3. The remaining four tasks all use conventional web-based interfaces and are similar in complexity, so we randomly assigned them to Agent~1 and Agent~2 while ensuring that each handled one everyday task and one office workflow. This design gives us coverage of three distinct UI patterns while keeping task--agent pairing a nuisance factor rather than a primary variable of interest.

\subsubsection{Procedure}
We recruited eight participants for a 45-minute user study, structured in two phases. Each participant completed all five tasks, allowing us to compare how both task type and confirmation frequency influenced user behavior within the same individuals.

Phase 1 — RQ1 (User Behavior): In Task 1, participants were instructed to check every step to familiarize themselves with the agent. In Tasks 2–5, they confirmed results only at the end of task execution. Participants were asked to think aloud, verbalizing their judgments about correctness and describing how they located errors in the sequence of steps. Phase 2 — RQ2 (Confirmation Preferences): We conducted a semi-structured interview to explore participants’ desired confirmation frequency, comparing their experiences with end-of-task versus step-by-step confirmation, and eliciting their ideal frequency. 

All sessions were recorded and transcribed. We conducted a thematic analysis \cite{inbook, beyer1999contextual}, in which two researchers collaboratively coded the transcripts. We refined emerging themes through the lens of our research questions (RQ1 \& 2).

\section{Formative Study Findings}
Our thematic analysis of the formative study data identified a dominant Confirmation–Diagnosis–Correction–Redo (CDCR) pattern where participants dealt with AI agent errors by reviewing the execution history from beginning to end. This strategy was evident across diverse task types and UI designs. We also observed some exceptions to this pattern (RQ1). For RQ2, participants expressed a desire for confirmation strategies beyond the prevalent ``confirm-at-end'' norm, suggesting opportunities for intermediate checkpoints. These insights informed the framing of our confirmation frequency model.

\subsection{Confirm-at-end is not enough}
Seven out of eight participants expressed their dissatisfaction with the current confirm-at-end strategy. As P4 noted:
\begin{quote}
``I did not like confirming at the end. Because the results were so far off from what I originally intended, that was a waste of time to do. Looking through the trace to try and find the error, I see that it was having a lot of other problems. If it had stopped there, maybe I could have clarified what my intentions were when I said the thing. Instead of wasting its time and getting something that was so far off from what I actually wanted to get.  And if we could maybe confirm at least a little bit more frequently, I think that would really help my confidence in the agent, [and improve] my communication [with the agent].''
\end{quote}
These findings suggest the dominant ``confirm-at-the-end'' approach should be reconsidered. Although Agent 2 \cite{openai2025agent} offers a reactive control mode (allowing users to seize control of the agent's screen), six of our eight participants reported that reactive control still made them uncertain about when to check, failing to help them “spend less time monitoring” (P1). 
Current agent UIs left users in a passive ``wait state'' during execution, with no meaningful hooks to monitor or steer the run. Consequently, to catch errors in real time, users are forced to maintain constant vigilance; if they look away, errors are easily missed. In that case, users cannot disengage while the agent executes, which undermines the value of automation. As P6 noted, ``If I have to check every step my agent is making, I'd rather do it myself.'' These results highlight the need for a proactive, agent-initiated confirmation mechanism that reduces users’ monitoring burden. The central challenge is to strike a balance between the fixed time cost of reviewing every agent action and the expected time cost of needing to confirm task completion at the end and then possibly undertake a long post-hoc diagnostic and redo process if the agent failed. We discuss our model for balancing between these two extremes with checkpoint scheduling in the next section. 

Scheduling confirmations requires consideration of two components: (1) the time costs for each part of the Confirmation-Diagnosis-Correction-Redo (CDCR) cycle, including the time associated with waiting for the agent to generate actions, and (2) the agent's accuracy. We found that when accuracy was high, users were fine with less frequent confirmation. 

\subsection{Common Usage Patterns}
\subsubsection{Common Agent Errors} 
The most frequent agent error in our formative study was a mismatch between the activity described by the agent (i.e., the action listed) and what was actually executed (i.e., the screenshot). For example, in the travel planning task, the activity list indicated ``click July 4th''as the start date, but the screenshot showed July 9th selected. Another frequent issue was that the agent did not always execute actions in full alignment with the user’s intended goal.

\subsubsection{The CDCR Pattern}
We identified a recurring Confirmation-Diagnosis-Correction-Redo (CDCR) sequence (Figure \ref{fig:CDCR}). After the agent completes a task, participants typically \textit{confirm} that the final screenshot matches their expected result. If not, they first \textit{diagnose} the error by scanning the entire execution history, often skimming the activity list or sliding through the screenshot recording. Once they get the whole picture of the execution process, they scan the execution sequence step-by-step from the beginning, reviewing the activity descriptions, reasoning traces, and associated screenshots along the way. After locating the erroneous step, participants usually \textit{correct} the agent by telling it how to fix the mistake. In the final \textit{redo} stage, participants wait for the agent to replan from the erroneous step to the end based on their correction.

\subsubsection{Exceptions to the CDCR pattern}
Across all 40 trials (5 tasks $\times$ 8 participants), 100\% (40/40) of trials entered the Confirmation stage, 87.5\% (35/40) proceeded to Diagnosis, and 82.5\% (33/40) executed a Correction. Agent differences did not fundamentally change this behavior. However, variations in task type and user familiarity revealed several edge cases in how the later CDCR stages were instantiated. (1) Ambiguous confirmability: In 5 trials (2 hotel-booking and 3 Overcooked), participants failed to identify an existing error. In the hotel-booking domain, ambiguity in what counted as “wrong” made it difficult to diagnose issues: for example, the agent sometimes chose a hotel that was slightly more expensive but had better transportation access. While arguably suboptimal, this choice was not objectively wrong. In the three Overcooked trials, participants noticed the unexpected result and entered the diagnosis phase, but failed to identify the first error. Because the agent continued to act after the initial mistake, participants reported that they expected the agent to correct itself later, and therefore overlooked the original error location. (2) Non-linear diagnosis: Two users skipped intermediate steps during diagnosis in Overcooked. Because they are highly experienced with Overcooked, they jumped directly to key states: after noticing the player placing soup on the wrong pad. (3) Partial correction: In 2 document-processing trials, participants entered the confirmation and diagnosis stages but chose an alternative correction path. They preferred to download the result file and fix the error themselves rather than having the agent re-execute the task, because the errors were minor and unlikely to affect subsequent planning.

\subsection{Implications for Model Scope and Applicability}
Edge cases in the previous subsection provide concrete examples of how users actually confirm, diagnose, and correct agent errors in practice. Building on these observations, we abstract a set of modeling assumptions that capture a common regime while making clear where the model does not apply.

\begin{enumerate}
    \item The task has a clear distinction between correct and incorrect outputs.
    \item The user is not deeply familiar with the task, and thus needs to review from the beginning to understand the agent’s actions. 
    \item Errors can propagate, and without timely correction, they affect subsequent similar actions.
    \item Once the current step is verified, the model treats all earlier steps as correct and only searches for errors in later steps.
\end{enumerate}
While factors such as task structure and user preference are important for real-world deployment, our current modeling abstracts away these user-specific and task-structural influences to focus on the core trade-off between confirmation cost and error diagnosis cost. This simplification positions our model within a broader landscape of supervision patterns, while leaving these additional dimensions for further investigation.

\begin{figure*}
    \centering
    \includegraphics[width=1\linewidth]{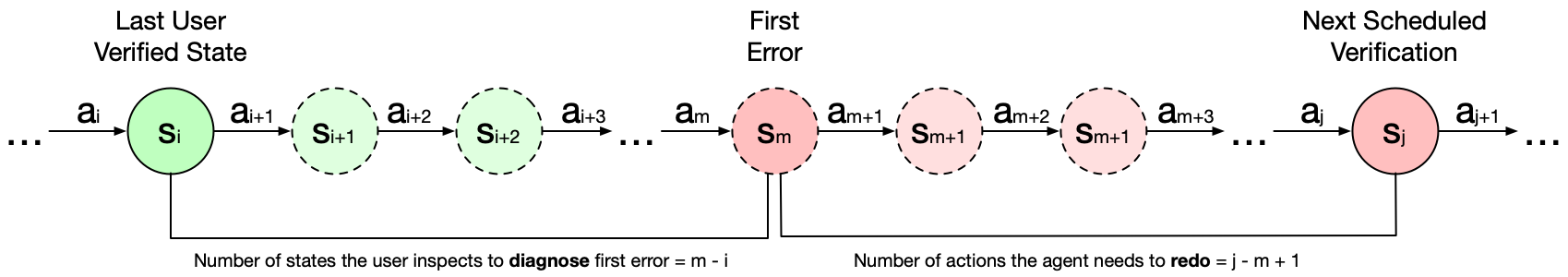}
    \caption{An example of the interval between the last verified state and the next confirmation point. The colors denote the correctness of the plan execution at that point, but this correctness is only observed when the user inspects them (denoted by solid circles). When the user finds an error in state $s_j$, they must go back to $s_i$ and check each state one at a time until they find the first error. This requires them to inspect $m-1$ states. Once the first error is identified, the agent then has to redo $j-m+1$ actions. Note, the dashed circles signify that the correctness of a state is not known until diagnosis.}
    \label{fig:example-interval}
\end{figure*}

Our intended application scenarios involve long, multi-step tasks with repeated sub-actions where each sub-action carries a non-trivial probability of error. For example, in image editing workflows involving repeated operations such as removing objects, changing backgrounds, and cropping images, each step is similar in structure but may still fail unpredictably. Our model aims to identify confirmation strategies that maximize efficiency in such contexts.

\section{Modeling Confirmation as a Minimum Time Scheduling Problem}

This section introduces a formal model of confirmation scheduling. Our goal is to determine \emph{when} the system should prompt the user to confirm correctness during a multi-step agentic task so as to minimize the user time needed to execute a whole task correctly. In other words, at each step, our system decides whether the agent should proceed autonomously or whether the user should be asked to step in and supervise. We proceed in three stages: (i) we define the environment (states, actions, transitions); (ii) we derive the expected time of a single confirmation interval using the \textsc{CDCR} decomposition; and (iii) we present a dynamic programming algorithm to identify the minimum time checkpoints. Finally, we describe how our model's parameters can be personalized.

\subsection{Problem Formulation}
We can formulate the scheduling of confirmations as a stochastic decision process:
\[
\mathcal{P} = (\mathcal{S}, \mathcal{A}, \mathcal{T}),
\]
where $\mathcal{S}$ is the state space, $\mathcal{A}$ is the action space, and $\mathcal{T}$ is the transition function. 

\subsubsection{State space $\mathcal{S}$} Given an agent action sequence of length N, there will be $2 \times (N+1)$ states in the space. Among these, $2N$ states correspond to the points after each action has been executed, which we denote as $s_i = \texttt{correct}$ or $s_i = \texttt{incorrect}$ for $i\in {1,\dots,N}$. The correctness value of $s_i$ is determined by whether all actions up to step $i$ have been executed correctly by the agent.
For example, if the system is in state $s_3=\texttt{correct}$, this means that the agent's first three actions have been executed correctly. In addition, there are two special states that are always correct: (1) an initial state $s_0$, representing the point before any actions have been executed, and (2) a terminal state $s_{\texttt{done}}$, indicating successful completion of the agent's entire action sequence.

At any given point, it knows that all agent actions up to the last user-verified state are correct, but it does not know whether the states beyond this point are correct (see Figure~\ref{fig:example-interval}).

\subsubsection{Action space $\mathcal{A}$.} In any given state $s_i$, the system can choose between one of two actions. It can either execute action $i+1$ from the agent's plan (\texttt{execute}) or request confirmation from the user (\texttt{verify}). Note that the system does not choose the agent's action (e.g., button clicks, scrolling, tool use), but only whether to proceed autonomously with execution or seek user verification. In this way, the system determines whether the agent or the user takes the lead. When the system is in state $s_0$ it can only choose to \texttt{execute} because this state is always correct by default, so confirmation is unnecessary. Similarly, when the system is in the final state $s_{N}$, it can only \texttt{verify} to transition to the terminal $s_{\texttt{done}}$ state because there are no more agent actions to execute.

\subsubsection{Transition function $\mathcal{T}$.}
The transition function depends on the two action options chosen by the system. We distinguish two cases:

\textit{(1) Execute.} When the system chooses to $\texttt{execute}$ it will transition from one state to another according to the following transition probabilities:
\begin{itemize}
    \item $p(s_{i}=\texttt{correct}|s_{i-1}=\texttt{correct}, \texttt{execute}) = p_{a_i}$
    \item $p(s_{i}=\texttt{incorrect}|s_{i-1}=\texttt{correct}, \texttt{execute}) = 1- p_{a_i}$
    \item $p(s_{i}=\texttt{correct}|s_{i-1}=\texttt{incorrect}, \texttt{execute}) = 0$
    \item $p(s_{i}=\texttt{incorrect}|s_{i-1}=\texttt{incorrect}, \texttt{execute}) = 1$
\end{itemize}
\noindent where $1 < i \leq N$ and $p_{a_i}$ corresponds to the probability that the agent will execute action $i$ from its plan correctly. These transitions also capture the cumulative probability of errors; i.e., once an agent makes a mistake, all subsequent states will be incorrect. Note that there is no \texttt{execute} transition out of the $s_N$ state. An important point here is that the correctness of all states after the last user-verified state is not directly observable to the system; it becomes known only when the user inspects them. To compute the probability that any given state $s_i$ is correct, the system must weigh the transition from the previous state by its beliefs about the correctness of the prior state: 
$p(s_i=\texttt{correct}) = p(s_{i-1}=\texttt{correct})p(s_i=\texttt{correct}|s_{i-1}=\texttt{correct}, \texttt{execute})$. By chaining over transitions, we can compute the probability that any state $j$ is correct given the last user-verified state $i$, where $i < j$: $p(s_j=\texttt{correct}|s_i=\texttt{correct}) = \prod_{m=i+1}^j p_{a_m}$. Note, we do not need to consider actions before the last user-verified step because we already know they are all correct.

\textit{(2) Verify.} When the system chooses $\texttt{verify}$, the user inspects the current state $s_j$ and determines if it is \texttt{correct}.  
\begin{itemize}
    \item If $s_j=\texttt{correct}$ and $j<N$, then $p(s_m=\texttt{correct})$ is set to 1 for all $m < j$ and system remains in $s_j$.
    \item If $s_j=\texttt{correct}$ and $j=N$, the system transitions to the terminal state $s_{\texttt{done}}$.
    \item If $s_j=\texttt{incorrect}$, then the user goes back to the last verified state and starts inspecting each of the following states to diagnose where the first error state occurs. If the last state verified by the user was $s_i=\texttt{correct}$ and during diagnosis the user determines that the first incorrect state is $s_m=\texttt{incorrect}$, then the system sets $p(s_k=\texttt{correct}) = 1$ for all $i<k<m$ and transitions to state $s_{m-1}$ (e.g., see Figure \ref{fig:example-interval}). The probability that the first incorrect state will occur at state $s_m$ follows a geometric distribution:     $p(\text{first \texttt{incorrect} at }s_m|s_i=\texttt{correct}) = 
    p(s_{i+1}=\texttt{correct}, \cdots, s_{m-1}=\texttt{correct}, s_m=\texttt{incorrect})
    = \left(\prod_{k=i+1}^{m-1} p_{a_k}\right)\,(1-p_{a_m})$.
\end{itemize}

\subsection{Time Cost}
\begin{figure*}
    \centering
    \includegraphics[width=1\linewidth]{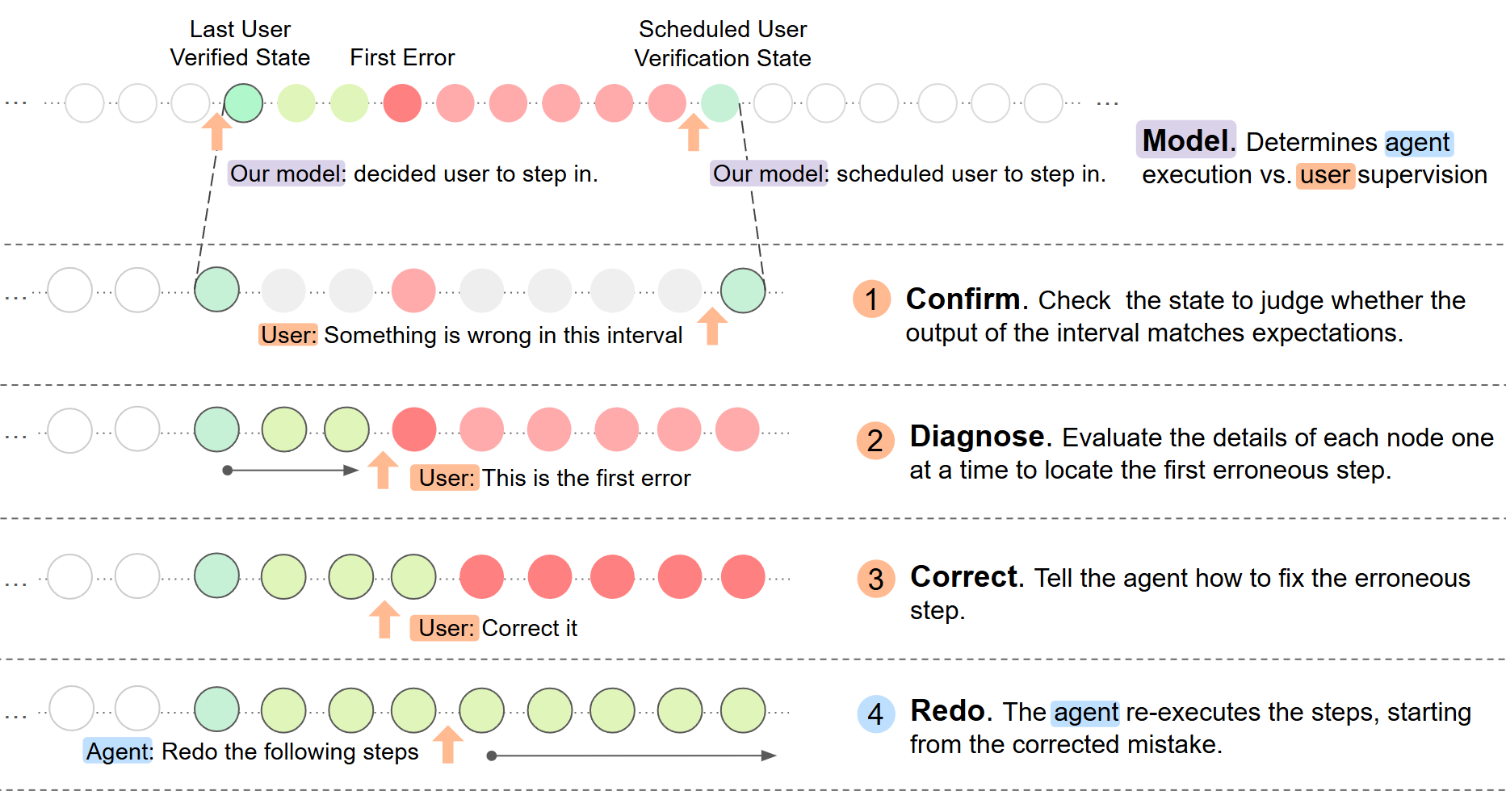}
    \caption{The Confirmation–Diagnosis–Correction–Redo (CDCR) interaction pattern and how our model schedules user verification.
Given a multi-step agent plan, the model selects when the user should step in (purple labels). At each scheduled checkpoint, the user (1) Confirms whether the interval’s output matches expectations; if something looks wrong, the user (2) Diagnoses the interval by inspecting nodes one by one to locate the first erroneous step; the user then (3) Corrects the error; and finally the agent (4) Redoes the remaining steps from the corrected point.}
    \label{fig:diag}   
\end{figure*}
The system measures cost in terms of \textit{user interaction time} (seconds), 
following the Confirmation--Diagnosis--Correction--Redo (CDCR) pattern, as shown in figure ~\ref{fig:diag}. 
We first derive the expected time cost from step $i$ to its next confirmation point $j$. The interval begins at a state $s_i=\texttt{correct}$, which corresponds to the last state that the user has inspected and determined to be correct. The system then repeatedly chooses to
\texttt{execute} actions until it reaches state $s_j$. At this point, the system chooses to ask the user to \texttt{verify} the correctness of $s_j$.

First, we denote the per-step time components in the CDCR pattern as follows:
\begin{itemize}
\item $t_{\text{confirm}}^i$: time for the user to confirm whether state $s_i$ is correct. 
  Concretely, the user compares the screenshot of the state with the agent’s activity list 
  (i.e., the sequence of executed actions) to verify correctness.

\item $t_{\text{diagnose}}^m$: the time required to inspect each intermediate state $s_m$ by looking at its screenshot and the description of the single agent action executed immediately before.\footnote{This is different from $t^i_{\text{confirm}}$, which requires evaluating a sequence of agent actions rather than a single agent action.}

  \item $t_{\text{correct}}^m$: time for the user to specify what went wrong with action $a_m$ and provide updated instructions to the agent to correct the mistake.

  \item $t_{\text{redo}}^m$: time for the agent to re-execute the action $a_m$ after the correction.
\end{itemize}

If we know that the user has verified all the states up to $s_i$, then we can compute the expected time to go from $i$ to $N$ if the next checkpoint is at $j$, where $i<j \leq N$, as:

\begin{align}
T[i,j] &= t_{\text{confirm}}^{j}
+ p(s_j=\texttt{correct}\mid s_i=\texttt{correct}) \min_{k} T[j,k] \notag\\
&\quad + \sum_{m=i+1}^{j} p(\text{first \texttt{incorrect} at } s_m\mid s_i=\texttt{correct})\notag\\
&\quad
\Bigl(
  \sum_{k=i+1}^{m} t_{\text{diagnose}}^{k} + t_{\text{correct}}^{m}
  + \sum_{k=m}^{j} t_{\text{redo}}^{k} + \min_{k} T[m-1,k]
\Bigr)
\end{align}

The first term corresponds to the time it takes the user to confirm state $s_j$. The second term corresponds to the case where the state $s_j$ is correct, which occurs with probability $p(s_j=\texttt{correct}|s_i=\texttt{correct})$. If this happens, then the expected remaining time is recursively defined as the expected time needed to traverse from $j$ to $N$ when the system chooses the next checkpoint $k$ such that this time is minimized (denoted as $\min_k T[j,k]$). The third term corresponds to the cases where $s_j=\texttt{incorrect}$. If this happens, we sum over all possible locations of the first error, weighted by their probabilities. For each first incorrect state $s_m$, the expected remaining time is composed of four parts: the time needed to diagnose the error, the time needed to correct the mistake, the time needed to redo the incorrect steps, and the expected time needed to go from $m$ to $N$ if the system chooses the next checkpoint $k$ such that this time is minimized.

\subsection{Dynamic Programming Checkpoint Solver}
\begin{figure*}
    \centering
    \includegraphics[width=1\linewidth]{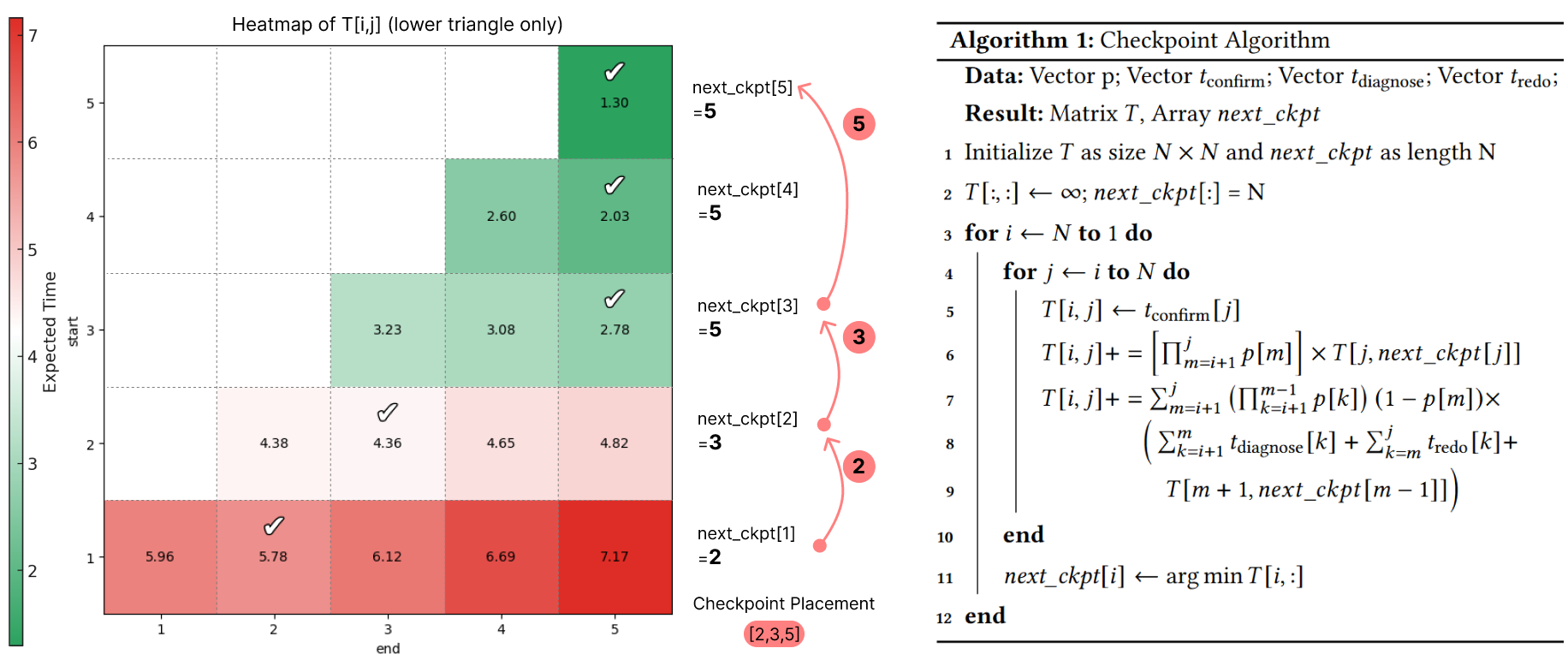}
    \caption{Dynamic programming algorithm and results for an agent action sequence of length $N=5$ with action success probabilities $p=[0.7,0.7,0.9,0.85,0.85]$. 
    Costs are parameterized as $t_{\text{confirm}}[k] = 1$, 
$t_{\text{diagnose}}[k] = 1$, and $t_{\text{redo}}[k] = 1$, 
for all $k \in \{0,\dots,N\}$.
    Each cell $[\textit{start}, \textit{end}]$ specifies the expected user time to correctly execute the rest of the action sequence from $start$ to $N$, if the system has a verification checkpoint at state \textit{end}. 
    White checkmarks indicate the $next\_ckpt$ locations for each $start$. The right-hand annotations summarize the resulting checkpoint policy.}
    \label{fig:dp}
\end{figure*}

Given the recurrent format of T[i, j], we developed a dynamic programming algorithm to efficiently identify the best verification locations that minimize the total user time, as shown in Figure ~\ref{fig:dp}. 

The algorithm starts at the end of the sequence, since there is always a verification at the end. It then works backwards, populating the values of $T[:,:]$ according to the previously defined formula.\footnote{We drop $t_{\text{correct}}^m$ (the time needed for the user to instruct the agent on how to fix $a_m$) from the dynamic programming update because all errors need to be fixed no matter when verifications happen, so it turns out that dropping this term does not change the checkpoint locations.}  At each iteration, the algorithm computes the best next checkpoint for each starting location $i$, denoted as $next\_ckpt[i]$.

The populated $next\_ckpt$ vector can be used to determine the best checkpoint locations. For a given starting location $i$ (which assumes $s_i=\texttt{correct}$), $next\_ckpt[i]$ specifies the location of the best next checkpoint. For the example in Figure \ref{fig:dp}, when starting at state 1, the best next checkpoint is at state 2, if starting at state 2, the best next checkpoint is at 3, and finally, when starting at state 3, the next best checkpoint is at state 5.

If the user determines that a state is incorrect during one of the checkpoint verifications, then they follow the CDCR procedure. This might result in them ending up in an earlier state. If the agent action sequence does not change, then the precomputed $T$ and $next\_ckpt$ can be reused without computation. If, however, the action sequence changes, then these need to be recomputed.

\subsection{Parameter Realism}
In our system, we have both agent-side (probability and redo time) and human-side parameters (confirmation and diagnosis time). The dynamic programming checkpoint solver itself is agnostic to how these parameters are obtained. Realistic values for these parameters can be calibrated from benchmark logs and user studies, and our algorithm permits these parameters to vary step-by-step, drift over time or update online, as illustrated via simulations (Appendix ~C).

\subsubsection{Probability}
On the agent side, evaluating agent performance is an active research area, with benchmarks reporting task accuracy \cite{xie2024osworld, valmeekam2023planbench, xie2024travelplanner, zhou2023webarena, xu2024theagentcompanybenchmarkingllmagents, ma2024spreadsheetbench}. From such traces, we can estimate per-step success probabilities $p_{a_i}$. Crucially, the model does not require a single constant $p$ across the entire task. We allow each step $i$ to have its own success probability $p_{a_i}$, which can depend on the tool, page template, or subtask. Appendix~C.1.1 shows how our model schedule checkpoints with heterogeneous step-wise accuracies.   

For practical deployments, where general benchmark data may differ from the user's specific environment, $p_{a_i}$ can be dynamically estimated using a Beta–Bernoulli belief model: $p_{a_i} \sim \mathrm{Beta}(\alpha_i, \beta_i)$. In this setup, accuracy metrics reported in benchmarks serve as informative priors. As environment conditions change (e.g., a website updates its layout or the user provides corrective feedback), the belief parameters can be updated online: incrementing $\alpha_i$ for autonomous successes and $\beta_i$ for failures. The posterior mean would then serve as the input to our dynamic program. While our core algorithm is agnostic to the source of $p_{a_i}$, this Bayesian approach demonstrates how the system could adapt to user-specific reality. Determining the optimal weight of the prior relative to new observations remains a nuanced trade-off for future exploration.

\subsubsection{Redo}
The execution time $t_{\text{redo}}^k$ can be grounded in timing benchmark results \cite{zhang2025optimizing, zhang2025leveraging, liu2023llm, waytowich2024atari}. Our model supports online replanning. When an error is confirmed at step $m$, we conceptually “cut” the original plan at that point: all steps up to $m-1$ are now treated as confirmed, and the remaining steps form a residual task of length $K'$ with a (possibly) updated action sequence. We then update the step-level parameters for this residual task (e.g., new success probabilities and redo times for the revised actions) and recompute the optimal checkpoint schedule for the remaining $K'$ steps using the same algorithm. In other words, the same dynamic program is simply re-invoked on the current residual plan, yielding a receding-horizon policy that naturally accommodates updated plans, changing hazards, and revised redo times without altering the core recursion.

\subsubsection{User Timing Parameter}
On the user side, our model includes two timing parameters: $t_{\text{confirm}}^k$ and $t_{\text{diagnose}}^k$. These are per-step quantities that may depend on interval-level or interface-level properties, such as the number of steps since the last checkpoint or the complexity of the UI (e.g., DOM size, density of interactive elements). Appendix~C.1.2 presents one concrete example. In practice, these parameters can be treated as functions of such features and estimated from pilot data or interaction logs. While different factors influence the \emph{numerical values} of the timing parameters—and therefore the resulting optimal schedule—they do not change the structure of our algorithm. A more detailed characterization of how specific interface factors shape these two parameters is an important direction for future work.

These two parameters $t_{\text{confirm}}^k$ and $t_{\text{diagnose}}^k$ can also be personalized. These quantities can capture hierarchical effects, with individual reaction times treated as adjustments to the population average and refined as more user data becomes available. In this paper, however, we focus on the population-level model and leave personalization to future work.

\section{Model Verification Study Design}
We compared two confirmation formats: intermediate confirmation, guided by our minimum time scheduling model, and confirm-at-end, the baseline in current AI agents. The study tested whether intermediate confirmation improves efficiency and user experience.

\subsection{Simulated Environment}
\begin{figure*}
    \centering
    \includegraphics[width=1\linewidth]{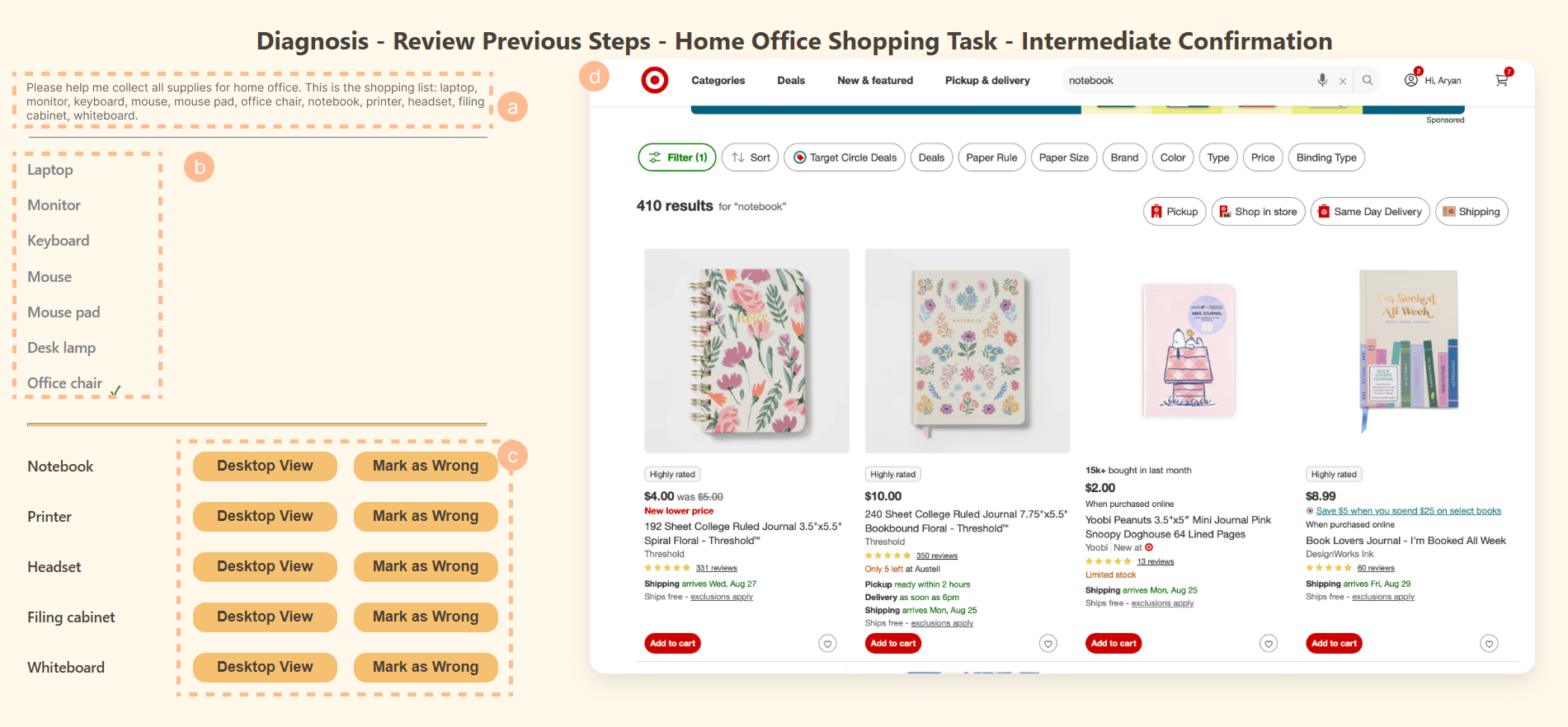}
    \caption{Diagnosis interface with clickable-node tracking for the Home Office Shopping task. (a) The user’s natural-language request to the agent. (b) Actions that the user has confirmed as correct. (c) The clickable-node tracking panel: each past action appears as a node that the user can click to open the corresponding agent execution state (“Desktop View”) or mark as the first wrong step (“Mark as Wrong”), which flags the error and triggers the agent to redo all subsequent actions. (d) The remote desktop view showing the agent’s execution state for the currently selected node.}
    \label{fig:env}
\end{figure*}

Real AI agents vary widely in execution time and outcome quality. We built a simulated environment that fixed execution time and accuracy for each task, allowing us to isolate confirmation frequency as the primary factor. The user interface used a clickable-node tracking design (shown in Figure~\ref{fig:env}), which participants in our formative study reported as helpful for navigation and understanding task progress. We selected three representative task domains: shopping cart management, image editing, and the Overcooked game. These cover three broad categories: daily life, work-related, and virtual tasks, while also differing substantially in execution time, tool usage, and workflow complexity. Error cases were drawn from real agent failures observed in prior experiments: in shopping, the agent failed to replace an out-of-stock item; in image editing, the agent applied an incorrect mask that prevented object removal; and in Overcooked, the agent grounded to the wrong ingredient dispenser (e.g., selecting the tomato instead of the onion). Appendix ~B shows details about the simulation environment.  

\subsection{Within-Subjects Design}
We employed a within-subjects design, where each participant completed two trials (intermediate and end) in all three domains. To control for potential biases, we introduced three layers of counterbalancing: task order, confirmation comparison order, and error location. 1) Task order. Each domain was treated as a block. We counterbalanced block order using a $3 \times 6$ Williams design, ensuring that each task domain appeared equally often in each position and that domain sequences were balanced across participants. 2) Confirmation strategy order. Within each block, the two confirmation formats for the same task were presented back-to-back. They were set to the same error location. To reduce familiarity effects, each trial followed the same structure but used different content (e.g., one shopping trial involved office supplies, another fitness items). Participants encountered different confirmation orders across the three blocks, and intermediate-first and end-first trials were balanced at the group level. 3) Error location. Error positions were sampled from early, middle, or late steps; shopping tasks included only early and late because these tasks were shorter. To prevent participants from anticipating that every trial would contain an error, we had each participant do one no-error trial. The no-error trial was implemented with fresh task content, assigned to one of the three task types, and counterbalanced across participants in both task type and confirmation format.

\subsection{Baseline Setting}
We selected confirm-at-the-end as our baseline because it is the dominant confirmation strategy in current agentic systems. As we observed in both system review and formative study in Section 3, current agents typically execute tasks autonomously from start to finish, and users only inspect the final result. Although some systems provide reactive controls, which are entirely user-initiated and offer no guidance on when the agent should proactively request confirmation. Because no deployed system offers such functionality, there is no established design to instantiate. In addition, introducing multiple hand-crafted policies would multiply study conditions without providing a principled comparison. For these reasons, we treat “confirm at the end” as the ecologically valid status quo. To complement the user study, we also provide simulations of several alternative confirmation strategies in Appendix ~D, allowing systematic comparison without burdening participants.

\subsection{Procedure}
The study began with a short tutorial explaining the agent background and confirmation methods. Each trial started with a task goal, and the Overcooked block included an additional introduction to the game rules. Each participant experienced 7 trials, including 2 (confirmation formats) * 3 (task types) + 1 no-error trial. After the trials, participants completed a post-study survey to capture participants' reflections on the confirmation strategies. The survey included open-ended questions pertaining to perceived task success, preferred confirmation style, and overall impressions of different confirmation strategies. It also included Likert-scale items assessing participants' perceptions of intermediate checkpoints (e.g., whether they saved time, reduced burden, or disrupted flow).

\subsection{Participants}
We recruited 48 participants from a university community (18 female, 30 male). Regarding LLM usage, 41 participants (85\%) reported using LLMs frequently (weekly or daily). In terms of familiarity with AI agents, $7$ (15\%) had never used them, $9$ (18\%) were beginners, $19$ (39\%) were intermediate users, and $13$ (28\%) were advanced users. Each session lasted about 15 minutes, and participants received \$5 compensation.

\subsection{Variable Determination} 
The confirmation frequency in our model is determined by four parameters: $t_{\text{confirm}}$, $t_{\text{diagnose}}$, $t_{\text{redo}}$, and the accuracy rate. Although the model is fully capable of supporting heterogeneous, per-step values, in our study, we instantiated these parameters at the per-domain level, reflecting the relative uniformity of actions within each task domain. For instance, shopping consists of repeatedly adding different items to a cart based on a predefined shopping list (e.g., a recipe list or office-supply checklist); image editing involves common editing operations such as object removal or color adjustment; and Overcooked involves primitive moves such as navigating to objects or placing items. Using fixed per-domain parameters avoids the high-variance “lucky vs. unlucky” trajectories that arise when per-step probabilities and timings are randomly varied, which would inject noise unrelated to our research question and obscure differences between confirmation strategies. Thus, instead of mixing heterogeneous values within a single trial and producing unstable or uninterpretable trajectories, we evaluate three domains with distinct parameter profiles to show that our model operates robustly across diverse settings.

Parameter values for the three domains were grounded in pilot studies and prior work.
(1) Confirmation and diagnosis times were estimated from a 10-participant pilot study with random checkpoints. In the main study, these averages were used as fixed inputs, with recalibration performed every 5–10 participants to prevent drift. We deliberately avoided per-user online adaptation, since stable personalization requires long-horizon usage data beyond the scope of this lab setting. (2) Execution time was estimated through formative agent runs and prior literature \cite{zhang2025optimizing, zhang2025leveraging, liu2023llm, waytowich2024atari}. Since steps are similar within each task, we applied fixed per-domain times: $t_{\text{redo}} = 20$s for shopping (dominated by network and screenshot analysis), $10$s for image editing (external tool calls), and $10$s for Overcooked (agent reasoning plus game state processing). (3) For accuracy and steps, we set fixed per-step action correctness probabilities for each task domain, derived from formative agent runs: 
$p_{\text{shopping}} = 87.5\%$ with 8 steps, 
$p_{\text{image editing}} = 91\%$ with 12 steps, and 
$p_{\text{Overcooked}} = 93\%$ with 16 steps. These values were applied uniformly across all steps within a task. To contextualize this parameter regime, a per-step success rate of 0.9 across a 10-step task yields roughly \(0.9^{10} \approx 0.35\) end-to-end accuracy, consistent with the \(\sim 30\%\) accuracy of recent multi-step benchmarks \cite{ma2024spreadsheetbench, xie2024osworld, xu2024theagentcompanybenchmarkingllmagents}.

\section{Model Verification Results}
In this section, we verify our model through both quantitative and qualitative analyses. While the quantitative results demonstrate what improvements intermediate confirmation achieves, the qualitative feedback explains why users value it and offers insights for future directions we discuss next.

\subsection{Quantitative Results}
\begin{figure*}
    \centering
    \includegraphics[width=1\linewidth]{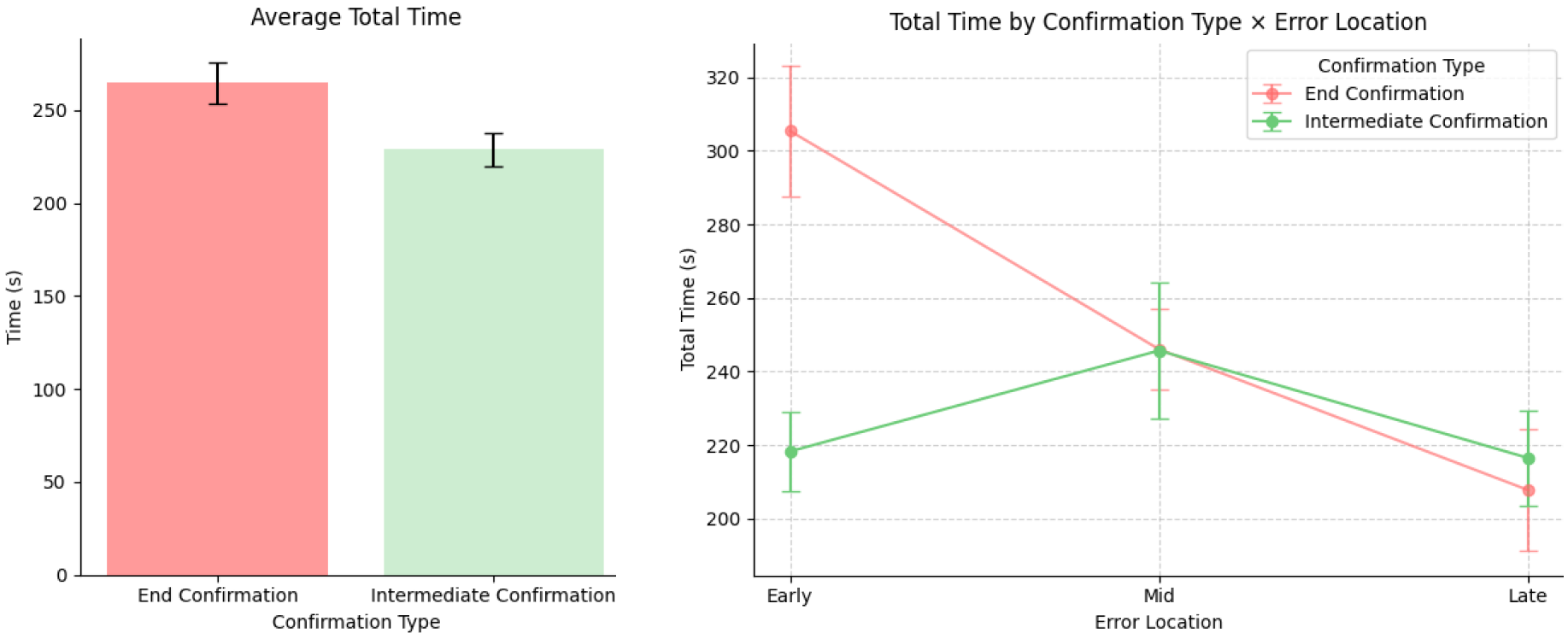}
    \caption{Average completion time under different confirmation strategies. Left: Intermediate confirmation reduced total time compared to end confirmation. Right: Effects vary by error location—large gains for early errors, minimal for mid, and slight cost for late errors. Error bars show 95\% CIs.}
    \label{fig:qual}
\end{figure*}

Across all task domains, our model reduced the completion time of a full task trial (8--16 steps) by \textbf{13.54\%}, corresponding to a savings of 35.84 seconds per task, compared to the baseline (End Confirmation) condition, $t(143) = 5.52, p < 0.001$. The baseline condition with end confirmation required an average of 264.70 seconds (95\% CI [253.43, 275.98]), while intermediate confirmation using our model required 228.86 seconds (95\% CI [219.73, 237.99]). Breaking down by task domain, the reductions were \textbf{17.44\%} (47.60s) for Shopping ($t(47) = 5.25, p < 0.001$; our model mean = 225.33s, 95\% CI [212.78, 237.88]), \textbf{7.46\%} (17.41s) for Image Editing ($t(47) = 2.34, p = 0.024$; our model mean = 215.85s, 95\% CI [203.42, 228.28]), and \textbf{15.64\%} (49.03s) for Overcooked ($t(40) = 2.14, p = 0.044$; our model mean = 264.51s, 95\% CI [236.71, 292.30]). 

The effect of confirmation frequency varied by error location. When errors occurred early, intermediate confirmation substantially reduced total time ($\approx 29\%$). For mid-task errors, the benefit was minimal ($\approx 2\%$), while for late errors, intermediate checkpoints slightly increased time ($+4.5\%$). This indicates that the value of intermediate confirmation is greatest for early error detection and diminishes as errors occur later in the task.

Across all tasks, the average completion time was $246.78$s (95\% CI [236.58, 256.98]), consisting of $13\%$ ($32.08$s, 95\% CI [28.29, 32.95]) confirmation, $9\%$ ($21.85$s, 95\% CI [16.99, 26.74]) diagnosis, and $78\%$ ($195.55$s, 95\% CI [189.30, 201.79]) execution. Shifting from end to intermediate confirmation meant spending more time on confirmations, but this was offset by large savings in diagnosis ($-17\%$) and redo ($-22\%$). Moreover, the average time per confirmation decreased by $38\%$, as users only needed to consider the most recent steps rather than reasoning over the entire task history when verifying correctness. Intermediate confirmation reduced the mental tax of each checkpoint. 

Beyond efficiency, reducing redo ($-22\%$) also cuts repeated LLM calls, lowering both financial costs and the energy footprint of unnecessary computation \cite{patterson2021carbon, luccioni2023estimating}. Although environmental impact is not our primary focus, this highlights an additional benefit: confirmation strategies that limit rework may also support more sustainable AI use.

\subsection{Qualitative Findings}

\subsubsection{Overall preference.} In our post-study survey, \textbf{81\%} of participants (39 of 48) preferred the modeled intermediate confirmation strategy, compared to 15\% (7 participants) who preferred end confirmation and 4\% (2 participants) with no preference (Fig.~\ref{fig:qual}, right). The Likert responses (Figure.~\ref{fig:qual}, left) further support this preference: most participants agreed that intermediate confirmation reduced errors (81\%), saved time (71\%), lowered cognitive burden (63\%), and helped prevent harmful errors (83\%). Notably, 77\% of participants disagreed that intermediate checkpoints disrupted their task flow.  As P18 noted, \textit{``I chose intermediate confirmation because it provides a balance between efficiency and reliability. By allowing the system to confirm at selected checkpoints, I can catch potential mistakes earlier instead of waiting until the very end, while still avoiding the constant interruptions of confirming every small step. This makes the process feel smoother but still safe.''} Overall, our modeled confirmation frequency achieved the intended goal of balancing user control with agent automation.

\subsubsection{Impact of Diagnosis.}Although diagnosis time accounted for only a small portion of total completion time in the quantitative study, participants emphasized that it had a strong influence on their experience. When errors were revealed only at the end, users reported feeling disoriented, as they had to review all steps at once without clear guidance: \textit{``If I can only confirm at the end, I feel confused about how to deal with the result, because I don’t know which step to check.''} (P39) Timely checkpoints minimized the effort needed for diagnosis, making error handling less confusing and less mentally taxing for users.

\subsubsection{Confirm at Early Stage.} The value of confirmations depends on the stage at which they occur. Based on our quantitative analysis, the largest time savings of intermediate confirmation arose when the agent made mistakes early in the task. The user survey further reinforced this finding from an experiential perspective, highlighting the importance of early-stage confirmation for both efficiency and usability. Each time a user and agent collaborate on a new task, the interaction is a dynamic process of gradually deepening mutual understanding: the agent gains a clearer understanding of the user’s needs, while the user gains a better sense of the agent’s performance and progressively builds trust in its automatic execution for that specific task. As P9 remarked:
\begin{quote}
``With an employee I haven't fully developed trust in yet, I would want to see intermediate steps in their work just to check that they've done it right. Perhaps if the employee has shown me that they can do the tasks correctly several times without errors, they can just show me their work at the end. I can't trust that it's going to do the task right all the way to the end quite yet, so I want it to show me that it has done the intermediate steps right before getting to a wrong final product.``   
\end{quote}
This perspective is also reflected in our model. We allow each step to have its own probability $p_{a_i}$, which can initially be set conservatively so that early-stage confirmations are more frequent. As users catch and correct early errors, $p_{a_i}$ increases, and the model dynamically decreases confirmation frequency. This adaptive behavior was not fully captured in our simulated environment. We will discuss in the next section how such dynamics could be more realistically incorporated.
\begin{figure*}
    \centering
    \includegraphics[width=1\linewidth]{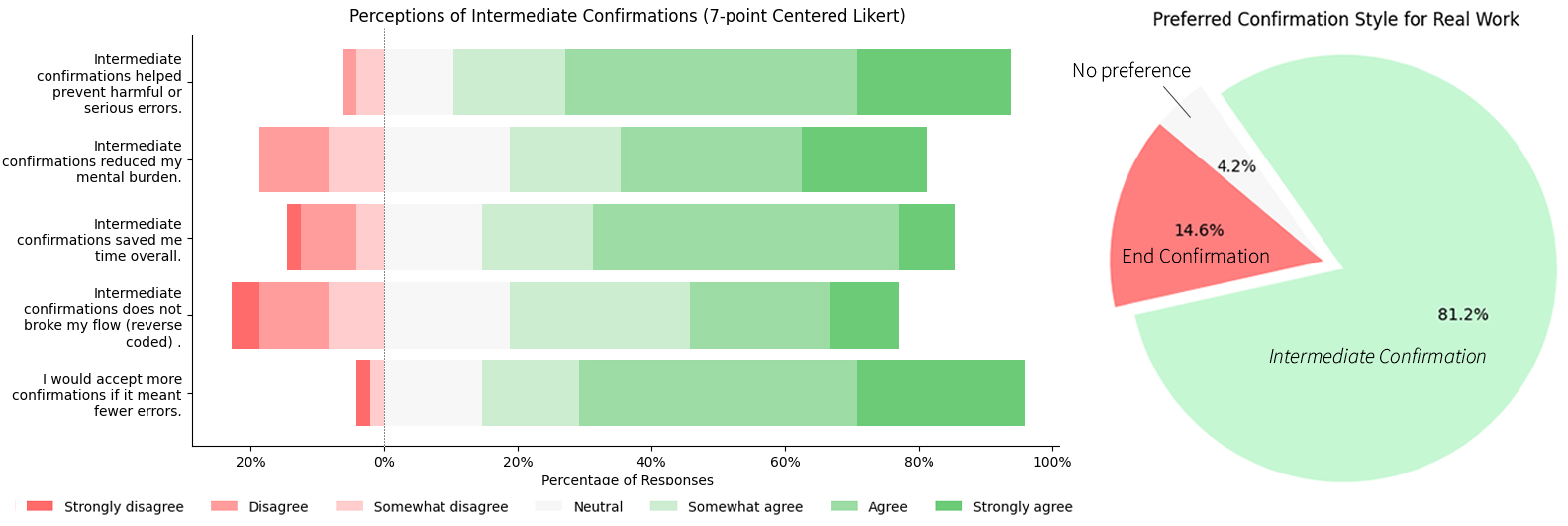}
    \caption{Participant Likert scale scores for the questions asked in our post-study survey (Left).Preferred confirmation style(Right)}
    \label{fig:qual}
\end{figure*}

\subsubsection{Subjective Perception.} Participants’ preferences for confirmation strategy also reflected their perceptions of task importance and difficulty. For important tasks, they preferred more frequent checks (e.g., ``something elaborate like trip planning”), while for trivial tasks (e.g., ``adding groceries to my cart”), they preferred confirmation only at the end (P20). Perceived difficulty also mattered: users expected easier tasks to require fewer confirmations. However, user perceptions often diverged from reality. For tasks they considered simple, participants expected fewer confirmations, yet were frustrated when the agent still made errors. For example, agents frequently failed in the shopping cart task, but many users saw this task as trivial. Since our model calculates confirmation frequency from actual error rates, it correctly prescribed more frequent checkpoints. While efficient in theory, such frequent confirmations may feel disruptive when they conflict with user expectations. In the next section, we discuss how confirmation strategies can be better aligned with user perceptions in future systems.

\section{Discussion}
Over the past decades, mixed-initiative systems have been moving from user-led tools \cite{li2025satori, arakawa2024prism, pu2025promemassist} toward increasingly agent-led workflows \cite{openai2025agent, anthropic2024claude}. As \citet{tennenhouse2000proactive} notes, humans shift from being “in the loop” on every step to acting as supervisors and policy-makers. Yet most current agents are still fundamentally reactive: they either wait for users to notice problems and intervene, or they only ask for input when they cannot proceed on their own. Our work aims at the next step in this trajectory: agents that remain mostly autonomous, but proactively allocate supervision by deciding when users should be brought back into the loop. Our scheduler can be plugged into existing systems as a timing layer that determines when to pause, ask for clarification, or surface plans for review. To move toward this vision, we consider two complementary directions: (1) what else to model—expanding beyond efficiency to capture subjective cost, safety, and trust, and (2) how else to design—reimagining confirmation as an interaction paradigm rather than a discrete interruption.

\subsection{Beyond Efficiency: Cost, Safety and Trust}
Our current model treats time as the only cost factor when determining the best confirmation points. However, in real-world applications, user decisions about when to confirm are influenced by broader considerations. As noted in Section 7.2.4, subjective perceptions play a critical role; here, we expand on this point by examining two key dimensions: real-world cost and user-perceived reliability.

\subsubsection{Real-world cost and safety} In practice, agentic failures can have an outsized real-world impact that extends far beyond execution time. Errors may lead to financial loss (e.g., rescheduling fees for a wrong flight), emotional or social consequences (e.g., accidentally sending an email to the wrong recipient), or even irreversible risks (e.g., permanently deleting a file) and harms (e.g., privacy breaches, opening phishing links, manipulative suggestions) \cite{alberts2024computers, bansal2024challenges, amodei2016concrete}. Strategic human oversight is therefore crucial for addressing these safety issues. Since human oversight is expensive, it cannot be applied uniformly to every low-level action; instead, it must be focused where it matters most \cite{holzinger2025human, amodei2016concrete}. Our CDCR framework speaks directly to this question of where oversight should be applied. Future work should expand the range of costs that can be supported. For general actions, the replanning term (e.g., $t_{redo})$ can be extended to support multi-dimensional cost functions that balance heterogeneous factors, such as time, money, emotional costs, and risks, replacing the purely time-based redo cost used in our current instantiation. Further, future work should also explore the use of mandatory checkpoints for high-stakes actions that occur regardless of the system’s estimated accuracy.

\subsubsection{Trust calibration across accuracy regimes}
Users’ willingness to confirm is shaped not only by actual system accuracy but also by how reliable they \emph{believe} the agent to be, based on subjective experience and agent framing \cite{lee2004trust, parasuraman2010complacency, shneiderman2020human}. Prior work shows that overly positive framings of system performance can inflate perceived reliability and encourage overuse even when this harms task outcomes \cite{cockburn2020framing, quinn2020loss, quinn2016bad}. Benchmark evaluations can exacerbate this gap: agents may appear highly capable on curated benchmarks while generalizing poorly to practical real-world tasks \cite{fodor2025line}. In such settings, optimistic benchmark-based messaging can lead users to over-trust the agent and reduce or even cease verification, reinforcing automation complacency. Our model offers a way to counter this by grounding the per-step correctness probabilities $p_{a_i}$ in observational real-world performance data (e.g., deployment logs or field studies) rather than benchmark scores alone. In that case, our model's recommended checkpoint schedule will be more conservative, supporting healthier user oversight and making benchmark optimism visible. As agent performance improves and true error rates drop, the role of user verification in our model correspondingly shifts: its value is less about catching frequent errors and more about maintaining calibrated trust, ensuring users retain occasional oversight instead of becoming fully complacent.

\subsection{Designing Confirmation for Transparency}
To better balance user control with agent automation, confirmation should be treated not only as a scheduling problem but also as an interaction design challenge. We focus on three aspects: goal alignment, information exposure, and integration with downstream agents for transparency.

\subsubsection{Plan alignment and control hooks} Our findings on error location highlight the importance of prevention. Errors introduced early often stem from poorly communicated goals rather than faulty execution \cite{bansal2024challenges}. While most HCI work emphasizes co-execution or error repair \cite{feng2024cocoa, arawjo2024chainforge, epperson2025interactive}, we identify a design opportunity in early-stage goal alignment. Before execution, agents could present a structured plan for user review, drawing on concepts from formal languages that use rigorous specifications to evaluate whether task goals are satisfied \cite{aeronautiques1998pddl, shivashankar2013hierarchical, nau2003shop2}. Recent agentic interfaces already move in this direction by exposing intermediate plans and approval guards to users, allowing them to correct goals, such as Magentic-UI and Ongoal \cite{mozannar2025magentic, coscia2025ongoal}. As noted in Section 7.2.3, our model assigns each agent action its own correctness probability $p_{a_i}$, which can be used to decide which parts of a proposed plan warrant mandatory review or additional clarification. In other words, these downstream agents define the interaction hooks for goal alignment, while our scheduler provides a way to time and target those hooks on the most consequential steps.

\subsubsection{Targeted transparency and information granularity} As tasks grow longer, participants in our study found confirmation increasingly difficult: extended linear logs made it hard to locate issues, leaving users disoriented. This highlights a core design question—what level of detail should agents expose for effective supervision? Current UIs that present step-by-step activity lists with screenshots are often too low-level, overwhelming users with detail without offering structure. Recent systems surface the same tension in different domains: some agents selectively pause at ambiguous decision points \cite{peng2025morae, feng2024cocoa, huq2025cowpilot}, some agents visualize execution traces in hierarchical way \cite{zhou2024improving}, and image agents maintain explicit intent states or reuse prior user preference models to reduce the number of refinement rounds \cite{hahnproactive, li2025efficient}. All of these systems must implicitly decide how much internal state and history to show, and when asking for additional input is worth the user’s effort. Agents should offer \emph{targeted transparency}, surfacing additional detail, explanations, or plan summaries only at checkpoints where the expected benefit of more information outweighs the cost of user attention. Hierarchical representations of agent behavior are one concrete instantiation of this idea: by grouping low-level actions into higher-level chunks, interfaces can let users quickly scan the structure of a run and then drill down only where something looks suspicious. If such hierarchical structuring is adopted, our CDCR pattern remains intact, but diagnosis time may decrease in a logarithmic fashion—since users could locate errors by checking higher-level chunks first (e.g., binary or ternary splits) rather than scanning all low-level steps. Our scheduler provides a normative basis for deciding where these targeted transparency checkpoints should sit. The same logic can be used to decide when to reveal more of the agent’s internal state as models become more accurate and routine errors become rare, shifting the role of confirmation from catching frequent mistakes to timing when additional transparency is most valuable.

\section{Conclusion}
This paper introduced a decision-theoretic model for determining when users should confirm agent actions in long-horizon tasks. Building on insights from a formative study that surfaced the recurring CDCR pattern and dissatisfaction with confirm-at-end strategies, we conducted a controlled experiment with 48 participants and found that intermediate confirmation reduced task completion time by 13.54\% and was strongly preferred by 81\% of participants. Looking ahead, our findings suggest that confirmation should be framed not as a binary trade-off between autonomy and oversight, but as a mixed-initiative design opportunity, pointing to future work on integrating efficiency, trust, and interaction design to create more reliable and user-supervised agentic systems.

\bibliographystyle{ACM-Reference-Format}
\bibliography{sources.bib}

\newpage
\appendix

\section{Formative Study Task Prompt}

\subsection{Task 1}
\noindent\fbox{%
    \parbox{\dimexpr\linewidth-2\fboxsep-2\fboxrule}{%
        Fill the cart on Kroger.com for a grocery delivery at 30332. I need all the ingredients to make a pizza at home. Maximize the quality of the ingredients while staying under the budget of 20 dollars.
    }%
}

\vspace{0.5em}
\noindent\textbf{Observed Errors: }The agent mentioned adding certain items to the cart but those items were not present at the end of the run. The final summary from the model did not match the contents of the cart.

\subsection{Task 2}
\noindent\fbox{%
    \parbox{\dimexpr\linewidth-2\fboxsep-2\fboxrule}{%
        Please help me edit this image. I want to create an advertisement for Coca Cola. Remove all the objects on the table except for the sprite. Change the sprite to coke. Remove the man in the background and change the background to an outdoor scene.
    }%
}

\vspace{0.5em}
\noindent\textbf{Observed Errors: }The agent often applied wrong filters, selected objects improperly. The final result was very far from the desired result.

\subsection{Task 3}
\noindent\fbox{%
    \parbox{\dimexpr\linewidth-2\fboxsep-2\fboxrule}{%
        Help me book a hotel in NYC for July 4 through July 12. It should be in Manhattan, clean, and within walking distance to several cafes. Look for good deals where the hotel is either less expensive than similar quality or on discount from its usual price.
    }%
}

\vspace{0.5em}
\noindent\textbf{Observed Errors: }The agent often selected the wrong dates for flight/hotel bookings.

\subsection{Task 4}
\noindent\fbox{%
    \parbox{\dimexpr\linewidth-2\fboxsep-2\fboxrule}{%
        Please turn these invoices into a spreadsheet which record each expense's time and cost. And then send this spreadsheet to my email.
    }%
}

\vspace{0.5em}
\noindent\textbf{Observed Errors: } There was a mismatch between the values provided in the invoice and the ones entered into the spreadsheet.

\subsection{Task 5}
\noindent\fbox{%
    \parbox{\dimexpr\linewidth-2\fboxsep-2\fboxrule}{%
        Please generate a step-by-step plan for cooking onion soup In the Overcooked game, there are two primitive actions: 1. go to: "goes to and faces the target object, where object is something like pot, onion, etc." 2. interact: "interact with the object, e.g., this should be called if you are trying to interact with the pot, onion dispenser, plate dispenser, tomato dispenser, tomato, onion, etc." 3. wait20: waits for 20 time steps The objects in the environment are pot, onion dispenser, plate dispenser, tomato dispenser, tomato, onion. Rules: 1. you should have onion soup before you cook an onion 2. every time you cook something, you have to wait 20 minutes
    }%
}

\vspace{0.5em}
\noindent\textbf{Observed Errors: } The agent selected the wrong location or did not execute the action it planned.

\section{Model Verification Study}
\subsection{Demographic Data}

\begin{table*}
\centering
{\small
\begin{tabular}{
  >{\raggedright\arraybackslash}p{0.35\textwidth}
  >{\raggedright\arraybackslash}p{0.6\textwidth}
}
\hline
\textbf{Demographic Information} & \textbf{Participant Counts} \\
\hline
AI Tool Usage Frequency & Never (1\%), Occasionally (13\%), Weekly (22\%), Daily (65\%) \\
Age & 18--25 (69\%), 26--35 (30\%), >55 (1\%) \\
Gender & Male (64\%), Female (36\%) \\
AI Tool Usage Proficiency & Unfamiliar (16\%), Beginner (18\%), Intermediate (39\%), Advanced (28\%) \\
\hline
\end{tabular}
\caption{Participant Information}
\Description{}
\label{tab:participant_information}
}
\end{table*}

\begin{figure}
    \centering
    \begin{subfigure}[b]{0.45\textwidth}
        \centering
        \includegraphics[width=\linewidth]{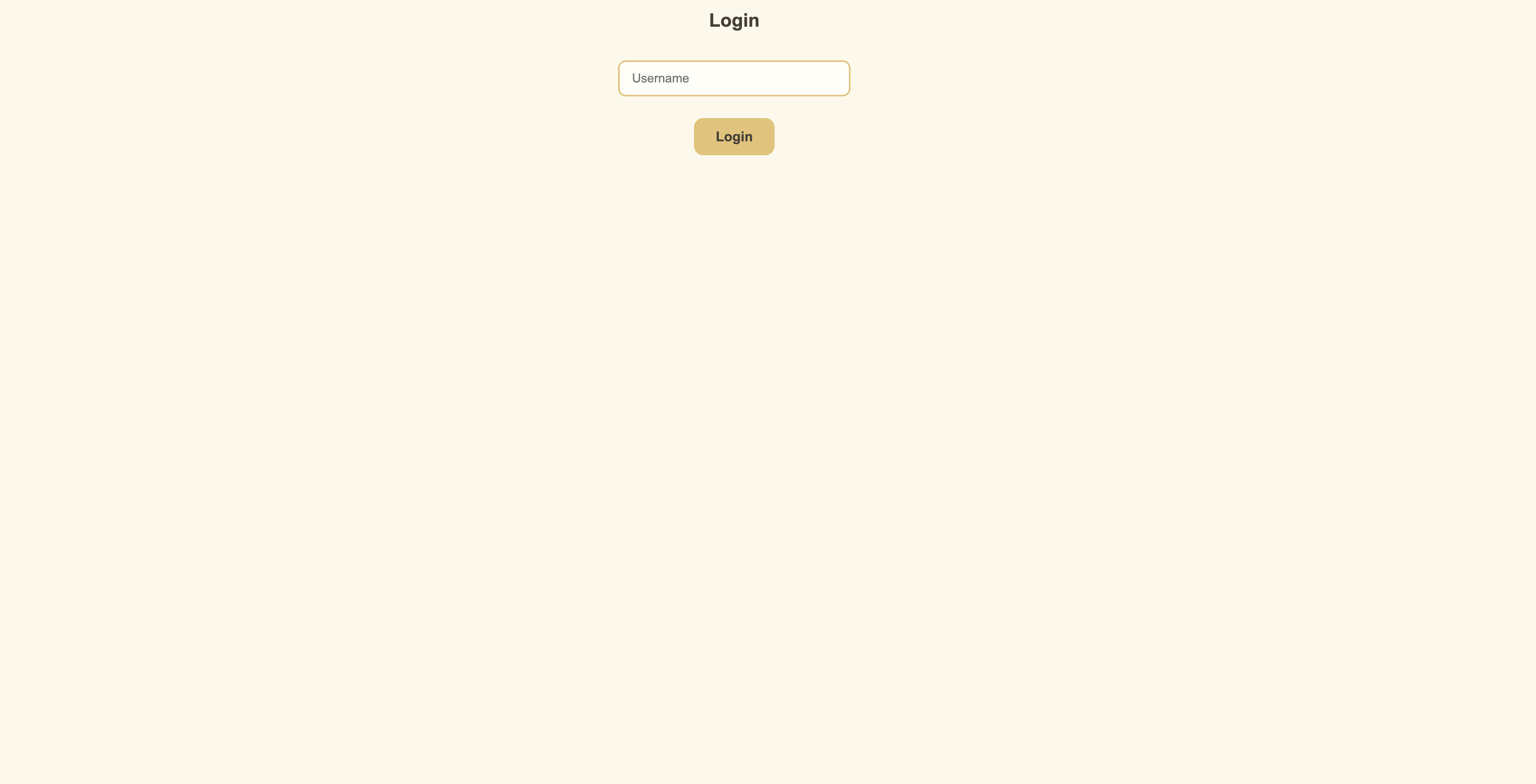}
        \caption{The Login Screen}
        \label{fig:login}
    \end{subfigure}
    \hfill 
    \begin{subfigure}[b]{0.45\textwidth}
        \centering
        \includegraphics[width=\linewidth]{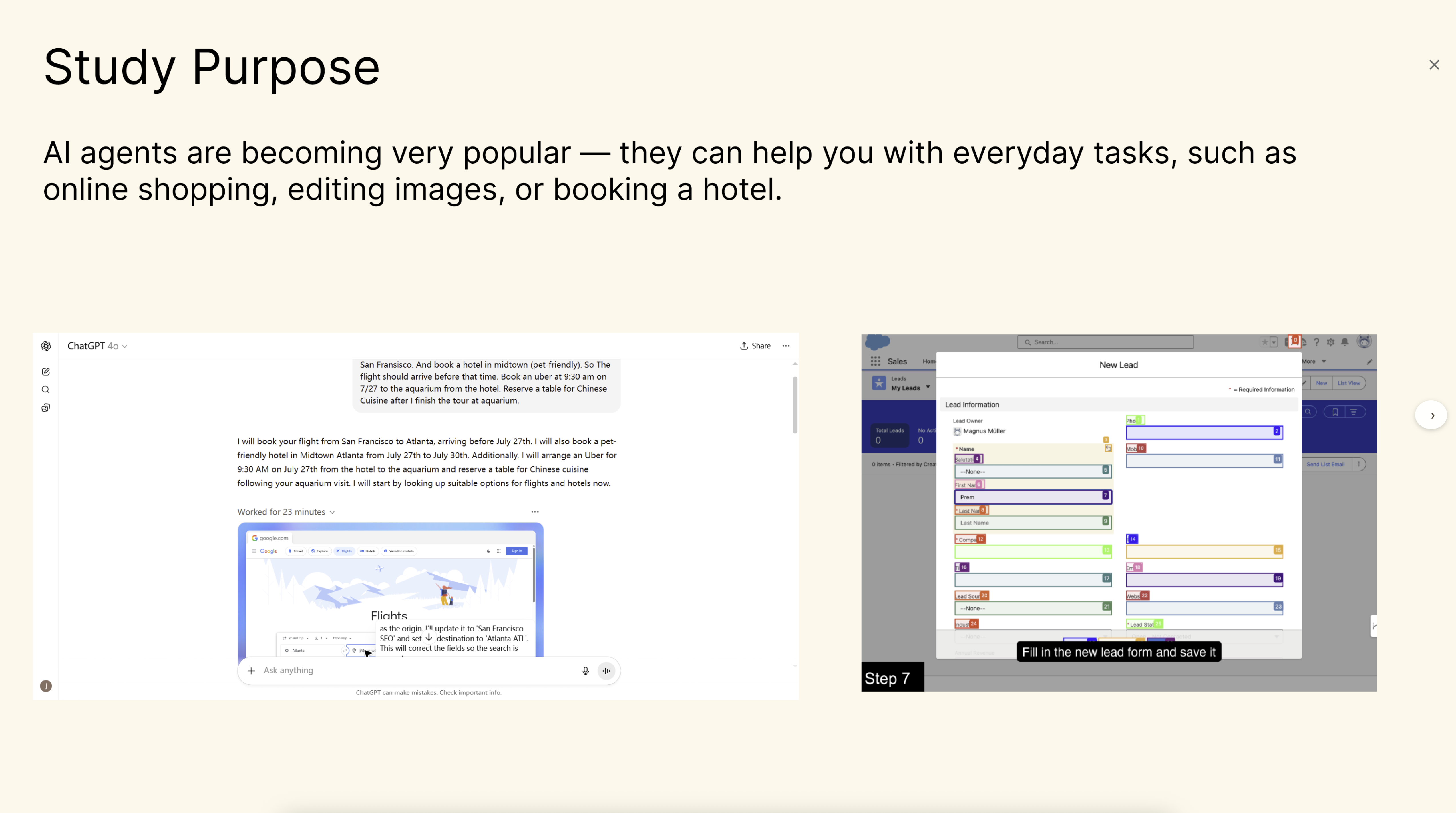}
        \caption{The Tutorial Screen.}
        \label{fig:tutorial}
    \end{subfigure}
    
    \caption{User Interface Introduction}
    \label{fig:ui_walkthrough}
\end{figure}

\subsection{User Study Design}

The user studies were conducted using a custom web application that presented participants with a simulated interface to an AI agent. The application was developed using standard HTML and CSS, with all experiment data logged to a Firebase Firestore database for subsequent analysis. The simulation comprised multiple screens that the user navigated sequentially to progress through the task. \\

\noindent \textbf{Tutorial and Login :} At the onset of the study, each user was presented with a short video tutorial within a modal popup, explaining the system's UI components. The tutorial demonstrated a person using the identical interface to monitor a simulated agent during a task. The introduction also included annotated images highlighting key UI elements, such as the agent state display and control buttons. After closing the tutorial, users were asked to enter their participant ID, which was assigned during recruitment. \\

\noindent \textbf{Task Execution:} Upon logging in, users were presented with a task summary and a description of the expected outcome. Data logging was initiated when the user clicked the ``Start Task'' button, transitioning them to the ``Working'' page. This page displayed a loading spinner and the text ``Agent is working'' to simulate the processing latency of a real AI agent between confirmation points. \\
\begin{figure}[htbp]
    \centering
    \begin{subfigure}[b]{0.45\textwidth}
        \centering
        \includegraphics[width=\linewidth]{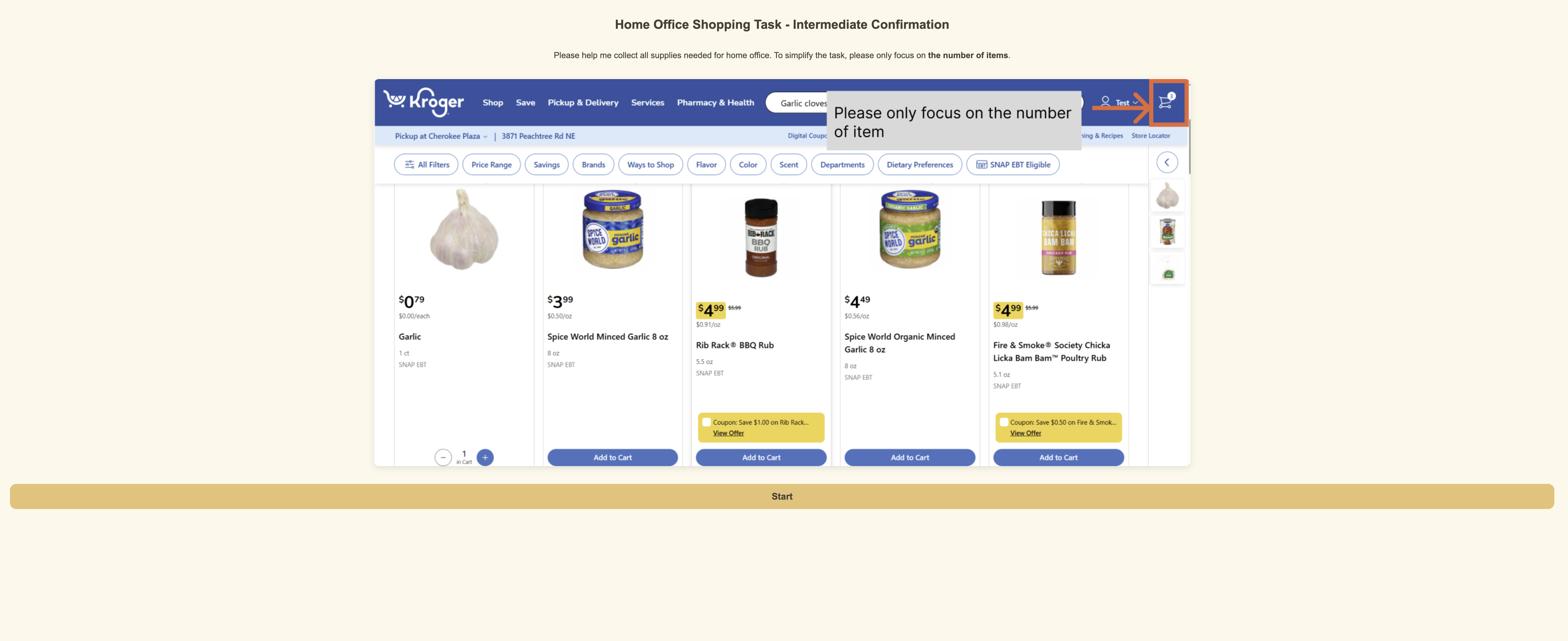}
        \caption{Task Start Screen}
        \label{fig:login}
    \end{subfigure}
    \hfill 
    \begin{subfigure}[b]{0.45\textwidth}
        \centering
        \includegraphics[width=\linewidth]{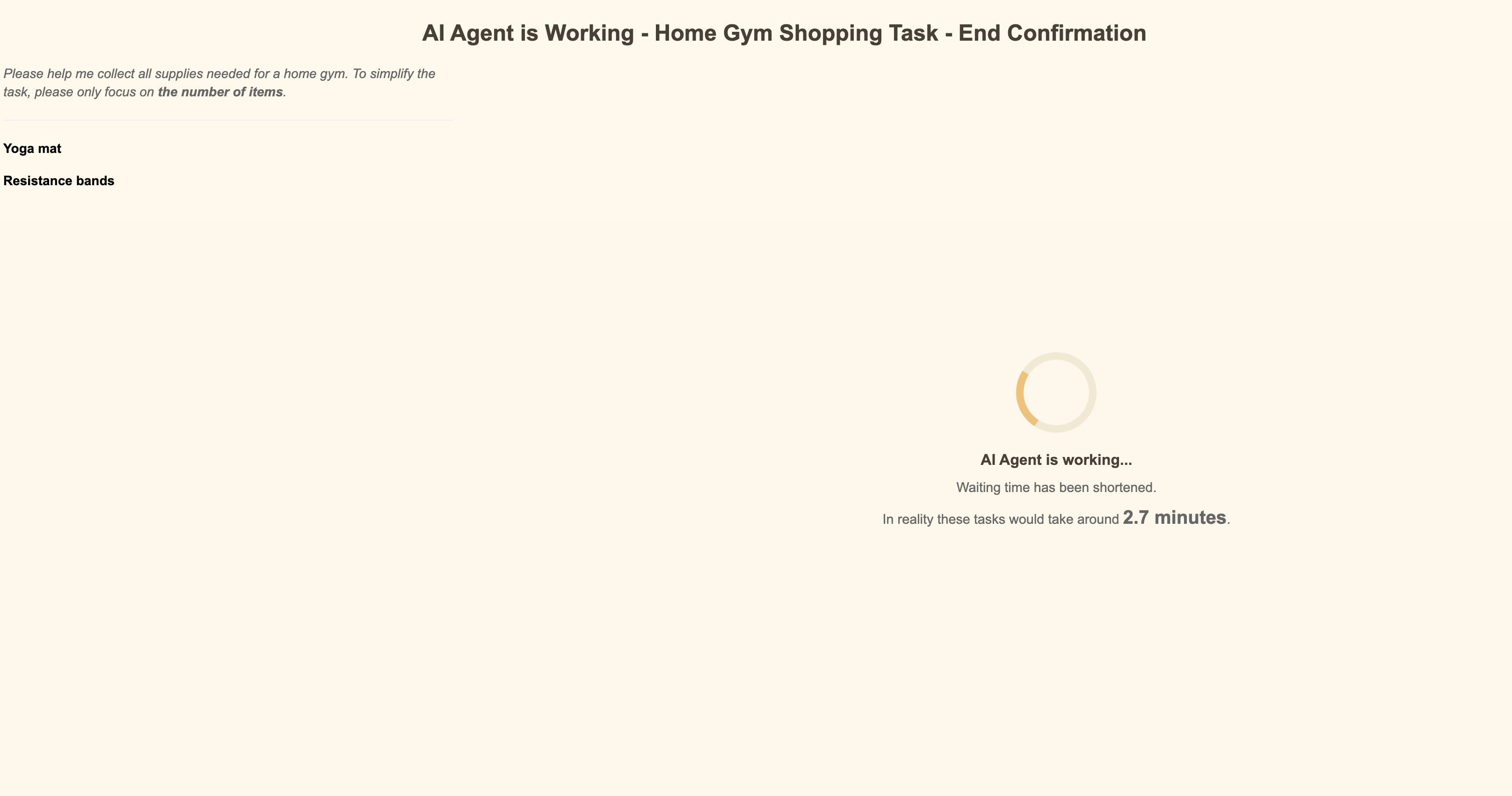}
        \caption{Working Screen}
        \label{fig:tutorial}
    \end{subfigure}
    
    \caption{Main task interface}
    \label{fig:ui_walkthrough}
\end{figure}

\noindent \textbf{Experiment Design :} Each participant ID was assigned a fixed sequence of experiments selected from a pool of image and tag pairs outlining the task flow. Each experiment was stored as a JSON object containing the step name, image reference, and a boolean value indicating correctness (used to determine which step contained an error). This sequence was structured to support the study's within-subjects design.

\subsection{User Study Task Description}

\textbf{Task 1: Shopping Cart Task :} In this task, the user was presented with a simulated AI agent tasked with purchasing a predetermined list of items from a home delivery platform. The user monitored the agent’s progress and was asked to identify any errors the agent committed whenever a checkpoint was reached.

\begin{itemize}
    \item \textbf{Subtasks:}
    \begin{enumerate}
        \item Type the name of the item in the search bar and initiate the search.
        \item Scroll through the results to identify the cheapest offering.
        \item Add the selected offering to the cart.
    \end{enumerate}
\end{itemize}
\textbf{Task 2: Image Editing Task :} In this task, the user monitored an AI agent responsible for editing an image to achieve a specific target output. The input images contained several objects that required removal or replacement to generate the final result.

\begin{itemize}
    \item \textbf{Subtasks:}
    \begin{enumerate}
        \item Locate the object of interest within the frame.
        \item Outline the object with high precision.
        \item Remove or replace the object.
        \item Repeat the steps above until the task is complete.
    \end{enumerate}
\end{itemize}
\textbf{Task 3: Overcooked Task :} This task involved monitoring an AI agent controlling a character within a simulated version of the game \textit{Overcooked}. The agent was tasked with adding a specific list of ingredients to a pot and periodically adjusting the temperature to correctly cook the dish.

\begin{itemize}
    \item \textbf{Subtasks:}
    \begin{enumerate}
        \item Navigate to the item dispenser.
        \item Retrieve the correct quantity of the specified item.
        \item Add the item to the pot and adjust the temperature.
        \item Repeat the previous steps for all listed ingredients.
    \end{enumerate}
\end{itemize}

\section{Simulation}

\subsubsection{Setup and Parameterization}
We instantiated the image-editing task using the same base parameters as in our main simulation: a 16-step task with $p=0.91$, $t_{\text{confirm}}=10$, $t_{\text{diagnose}}=3$, and $t_{\text{redo}}=10$. To demonstrate that our model naturally supports heterogeneous, per-step specifications, we introduced controlled step-wise variations on top of this base regime.
\subsection{Heterogeneous Step-Wise Parameters}
\subsubsection{Step-Wise Probabilities and an Error Cluster}
\begin{figure}
  \centering
  \includegraphics[width=1\linewidth]{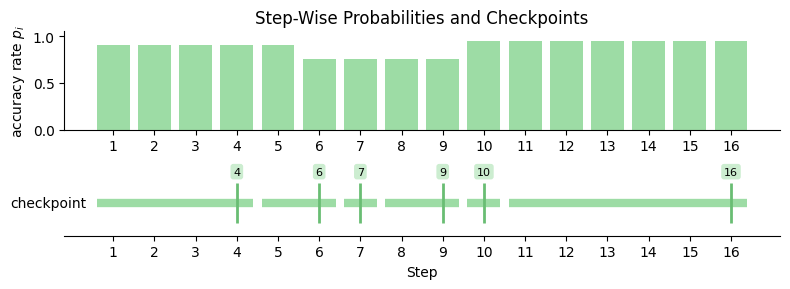}
  \caption{Step-wise correctness probabilities and the corresponding optimal checkpoint schedule}
  \Description{The first five steps have moderate reliability ($p=0.9$), steps 6--9 form an error-prone cluster ($p=0.75$), and steps 10--16 are highly reliable ($p=0.95$). The model allocates denser checkpoints within the brittle region (steps 4, 6, 7, 9, 10), and sparsifies them in the more reliable segments, adaptively concentrating user supervision where errors are most likely to arise.}
  \label{fig:p}
\end{figure}

We assigned non-uniform correctness probabilities across the 16 steps: the first five steps are moderately reliable ($p_{a_1}=\cdots=p_{a_5}=0.9$), the next four steps form a brittle error-prone cluster ($p_{a_6}=\cdots=p_{a_9}=0.75$), and the remaining seven steps are highly reliable ($p_{a_{10}}=\cdots=p_{a_{16}}=0.95$). As expected, the optimal checkpoint schedule increases its confirmation density within the brittle segment, indicating that the model adapts intervention frequency to localized spikes in error likelihood, as shown in Figure ~\ref{fig:p}.

\subsubsection{Step-Wise Timing Parameters}
\begin{figure}
  \centering
  \includegraphics[width=1\linewidth]{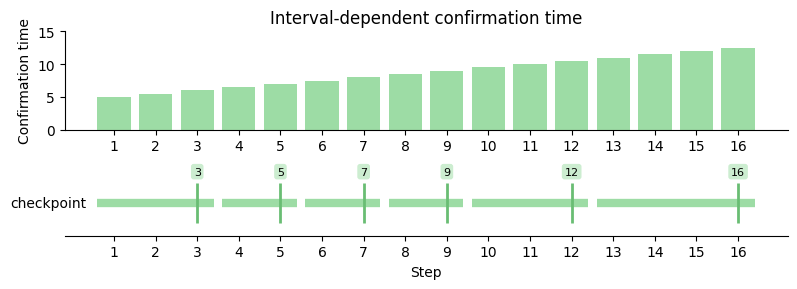}
  \caption{Interval-dependent confirmation time and the resulting optimal checkpoint schedule}
  \Description{ We set $t_{\text{confirm}}^k = 5 + 0.5(k-1)$, making confirmations progressively more expensive at later steps. The optimal policy correspondingly widens the spacing between checkpoints (steps 1, 3, 5, 7, 9, 12, 16), showing that the model shifts supervision toward earlier, cheaper intervals and avoids late confirmations unless necessary.}
  \label{fig:conf}
\end{figure}

To probe how heterogeneous timing affects the optimal schedule, we next made the confirmation time depend on the position of the step in the task. Specifically, for a 16-step task we set
\[
t_{\text{confirm}}^k = 5 + 0.5\,(k-1), \quad k = 1,\dots,16,
\]
so that confirming later steps becomes progressively more expensive. As a result, the optimal policy spreads checkpoints farther apart toward the end of the trajectory, trading off higher accumulated error risk against the rising marginal cost of confirmation, as shown in Figure ~\ref{fig:conf}.
We additionally simulated step-wise variants of diagnosis time $t_{\text{diagnose}}^k$ and execution/redo time $t_{\text{redo}}^k$. Across these heterogeneous timing regimes, we observed consistent qualitative trends: increasing confirmation time leads to wider spacing between checkpoints, while decreasing diagnosis or execution time makes it cheaper to discover and repair errors after they occur, and thus also increases the optimal span between checkpoints:
\[
t_{\text{confirm}} \uparrow \;\Rightarrow\; \text{span} \uparrow \qquad
t_{\text{diagnose}} \downarrow \;\Rightarrow\; \text{span} \uparrow \qquad
t_{\text{redo}} \downarrow \;\Rightarrow\; \text{span} \uparrow
\]

\subsection{Sensitivity to Per-Step Reliability}
\begin{figure}
  \centering
  \includegraphics[width=0.9\linewidth]{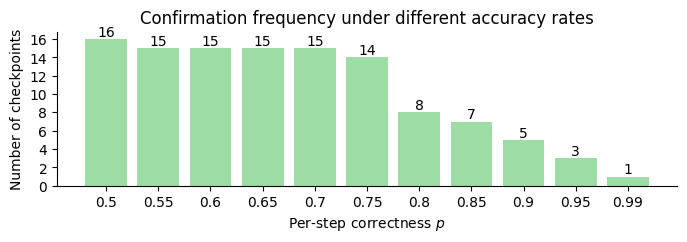}
  \caption{Sensitivity of the optimal checkpoint schedule to uniform per-step correctness in the image-editing domain.}
  \Description{A bar chart where the x-axis is per-step correctness and the y-axis is the number of checkpoints in the optimal schedule for a 16-step image-editing task. Bars are higher for lower $p$, showing that the scheduler responds to reduced reliability by increasing confirmation frequency.}
  \label{fig:sensitivity-p}
\end{figure}

We conducted a simple sensitivity analysis by sweeping the per-step correctness rate $p$ from 0.60 to 0.99 and computing the corresponding optimal checkpoint frequency. As shown in Fig.~\ref{fig:sensitivity-p}, in this 16-step task, the number of confirmation points decreases sharply as reliability increases. When $p \le 0.70$, the model places checkpoints at nearly every step, as the compounding error risk quickly overwhelms the cost of confirmation. Once accuracy exceeds $p=0.80$, checkpoint frequency begins to drop rapidly, and at near-perfect accuracy ($p=0.99$), the schedule collapses to a single confirmation at the end.

\section{Simulation on Probability-Based Baselines} 
\begin{figure}
  \centering
  \includegraphics[width=0.9\linewidth]{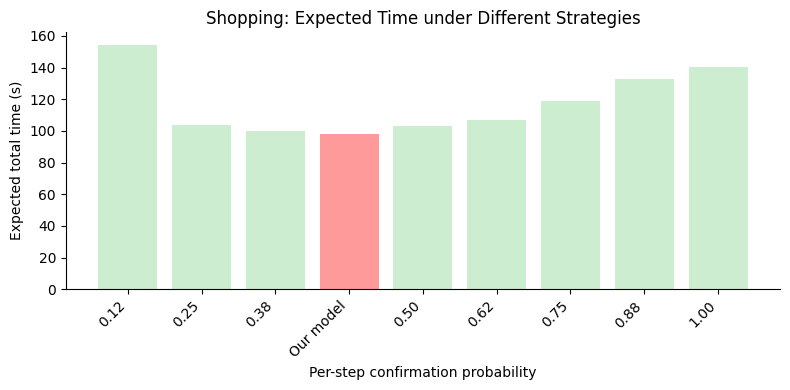}
  \caption{Shopping: Estimated completion time under probability-based confirmation strategies (green) versus our model’s scheduler (red). }
  \Description{Bar chart for the shopping domain with expected total time on the y-axis and per-step confirmation probability on the x-axis. Several light-green bars show different probability levels; the red “Our model” bar is close to the lowest green bar, illustrating that the model’s schedule is as good as or slightly better than the best probability-based strategy.}
  \label{fig:shopping}
\end{figure}

\begin{figure}
  \centering
  \includegraphics[width=0.9\linewidth]{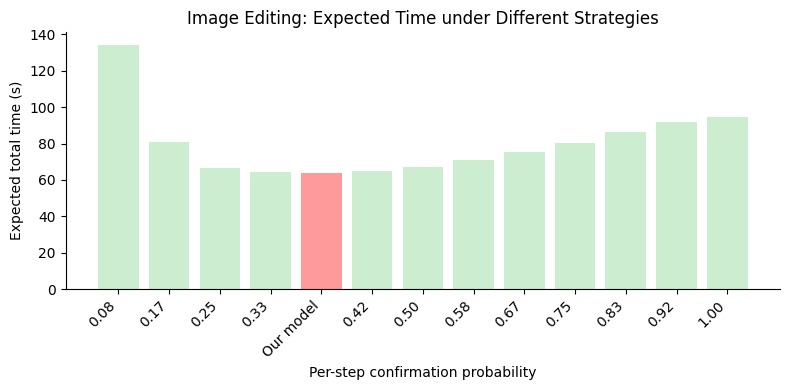}
  \caption{Image-editing: Estimated completion time under probability-based confirmation strategies (green) versus our model’s scheduler (red).}
  \Description{Bar chart for the image-editing domain with expected total time on the y-axis and per-step confirmation probability on the x-axis. Light-green bars show the expected time for different confirmation probabilities, forming a shallow U-shape. A red bar labeled “Our model” appears near the bottom of the U, indicating that the model’s scheduler performs as well as or better than any of the probability-based baselines.}
  \label{fig:image}
\end{figure}

\begin{figure}
  \centering
  \includegraphics[width=0.9\linewidth]{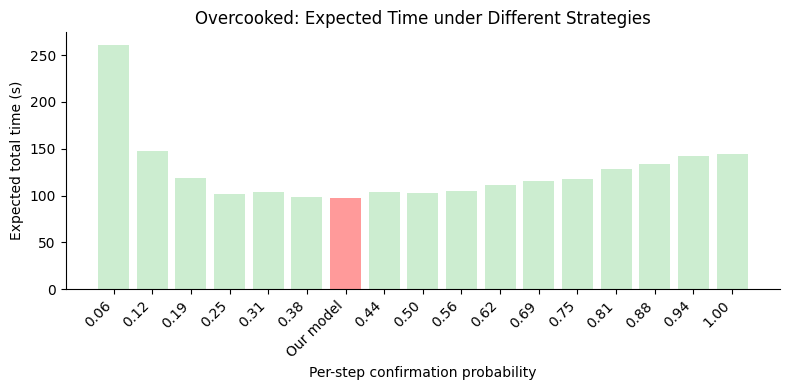}
  \caption{Overcooked: Estimated completion time under probability-based confirmation strategies (green) versus our model’s scheduler (red).}
  \Description{Bar chart for the Overcooked domain with expected total time on the y-axis and per-step confirmation probability on the x-axis. Light-green bars show the expected time for different confirmation probabilities, forming a shallow U-shape. A red bar labeled “Our model” appears near the bottom of the U, indicating that the model’s scheduler performs as well as or better than any of the probability-based baselines.}
  \label{fig:cook}
\end{figure}

As a complement to our user study, we ran a stochastic simulation that instantiates a family of probability-based confirmation policies and compares them against our model’s checkpoint schedule. In all three domains (shopping, image editing, and Overcooked), we reused exactly the same parameters as in the user study, including task length, per-step action accuracy, and the timing parameters for confirmation, diagnosis, and redo.

Under the probability-based policy, each step has an independent probability $p_{\text{check}}$ of becoming a checkpoint. For an $N$-step task, this yields an expected number of checkpoints of approximately $p_{\text{check}}*N$. To cover a wide range of behaviors, we swept $p_{\text{check}}$ over values whose expected number of checkpoints ranges from roughly 1 to $N$, thus spanning almost all reasonable confirmation frequencies.

For each schedule, we simulated 5,000 random runs. In each run, every step independently succeeded or failed according to the per-step accuracy, and we applied the CDCR pattern to compute the total completion time based on when checkpoints occurred and where the first error appeared. We then averaged the completion times over these 5,000 runs to obtain an empirical estimate of the expected time for that schedule.

Figures~\ref{fig:shopping}~\ref{fig:cook} plot the estimated completion time for all probability-based baselines, with an additional bar for our model.
Across all three domains, the probability-based policies exhibit the expected U-shaped pattern: very small $p_{\text{check}}$ values lead to few checkpoints and large rollback costs, while very large $p_{\text{check}}$ values approach confirming at every step and incur high confirmation overhead. Our model’s schedule (red bar) consistently lies the bottom of this curve and slightly lower than, the minimum across all probability baselines in each domain. Because our model uses the same number of checkpoints as the best-performing probability baseline, but optimizes their exact placement rather than only fixing the total count. 
\end{document}